%% file: paper.tex
\documentclass[sigconf]{acmart}

\makeatletter
\def\@ACM@checkaffil{
    \if@ACM@instpresent\else
    \ClassWarningNoLine{\@classname}{No institution present for an affiliation}%
    \fi
    \if@ACM@citypresent\else
    \ClassWarningNoLine{\@classname}{No city present for an affiliation}%
    \fi
    \if@ACM@countrypresent\else
        \ClassWarningNoLine{\@classname}{No country present for an affiliation}%
    \fi
}
\makeatother

\setcopyright{none}
\renewcommand\footnotetextcopyrightpermission[1]{} 

\DeclareTextCommandDefault{\nobreakspace}{\leavevmode\nobreak\ }

\usepackage{pifont}
\usepackage{pgfplotstable}
\usepackage{paralist}
\usepackage{filecontents}
\usepackage{amsmath}
\usepackage{multirow}
\usepackage{epsfig,endnotes}
\usepackage{tikz}
\usepackage{pgfplots}
\usepackage[ruled,vlined,linesnumbered]{algorithm2e}
\usepackage{amsfonts}
\usepackage{url}
\usepackage{tikz}
\usepackage{array}
\usepackage{booktabs,adjustbox}
\usepackage{etex}
\usepackage{color}
\usepackage{pstricks}
\usetikzlibrary{calc,positioning,shapes.geometric,shapes.symbols,shadows,arrows}
\usepackage{wasysym}
\usepackage{paralist}

\usepackage[justification=justified]{caption}
\usepackage{pgfplots}
\usetikzlibrary{arrows}
\usetikzlibrary{patterns}
 
\usepackage{listings}
\usepackage{multicol}
\usepackage[frozencache,cachedir=minted-cache]{minted}
\usepackage{lipsum}
\usepackage{adjustbox}
\usepackage{tabularx}
\usepackage{multirow}

\usepackage[english]{babel}
\usepackage{listings}
\usepackage{subcaption}
\usepackage{graphicx}
\usepackage{wrapfig}
\usepackage{hyperref}
\hypersetup{
    colorlinks = true,
    linkcolor = blue,
    anchorcolor = blue,
    citecolor = blue,
    filecolor = blue,
    urlcolor = blue
}
\usepackage{breakurl}
\usepackage{macro}
\usepackage[all]{nowidow}
\newcommand{\Xin}[1]{{\authnote{Xin}{#1}}}

\begin{document}
\date{}
\title{Understanding Binary Code Semantics with 
Large Language Models: How Far Yet?} 

\title{From Binary Code to Meaning: How Effective Are ChatGPT/GPT4 and Other 
Large Language Models?}

\title{From Binary Code to Semantic Understanding: \\ Evaluating the Effectiveness of ChatGPT/GPT-4 and Other Large Language Models}

\title{From Binary Code to Semantic Understanding: Benchmarking ChatGPT/GPT4 and Other Large Language Models}

\title{Binary Code Summarization: Benchmarking ChatGPT/GPT-4 and Other Large Language Models}


\author{Xin Jin}
\authornote{This work was partially completed during the author's internship at Microsoft Research.}
\affiliation{%
	\institution{The Ohio State University}
}
\email{jin.967@osu.edu}

\author{Jonathan Larson}
\affiliation{%
	\institution{Microsoft Research}
}
\email{jolarso@microsoft.com}

\author{Weiwei Yang}
\affiliation{%
	\institution{Microsoft Research}
}
\email{weiwei.yang@microsoft.com}

\author{Zhiqiang Lin}
\affiliation{%
	\institution{The Ohio State University}
}
\email{zlin@cse.ohio-state.edu}

\begin{abstract}
\input{sections/abstract}
\end{abstract}

\maketitle
\pagestyle{plain}

\section{Introduction}
\label{sec:intro}
\input{sections/introduction_new}



\section{Design}
\label{sec:design}

\input{sections/design}

 \section{Implementation}
\label{sec:implementation}

\input{sections/implementation}

\section{Evaluation}
\label{sec:eval}
 \input{sections/evaluation}


\section{Lessons Learnt}
\label{sec:lession}
\input{sections/lesson}

\section{Threats to Validity}
\label{sec:discussion}
\input{sections/discussion}

\section{Related Work}
\label{sec:related}
\input{sections/related}

\section{Conclusion}
\label{sec:conclusion}
\input{sections/conclusion}

{\footnotesize
\bibliographystyle{plain}
\bibliography{paper}
}

\remove{
\appendix
\section{Appendix}
\input{sections/appendix}

}

\end{document}

%% file: sections/abstract.tex
Binary code summarization, while invaluable for understanding code semantics, is challenging due to its labor-intensive nature. This study delves into the potential of large language models (LLMs) for binary code comprehension. To this end, we present \sysname, a comprehensive benchmark and dataset of over 557K binary functions and introduce a novel method for prompt synthesis and optimization. To more accurately gauge LLM performance, we also propose a new semantic similarity metric that surpasses traditional exact-match approaches. Our extensive evaluation of prominent LLMs, including ChatGPT, GPT-4, Llama 2, and Code Llama, reveals 10 pivotal insights. This evaluation generates 4 billion inference tokens, incurred a total expense of 11,418 US dollars and 873 NVIDIA A100 GPU hours. Our findings highlight both the transformative potential of LLMs in this field and the challenges yet to be overcome.

\ignore{
Binary code summarization is extremely helpful for comprehending ing binary code semantics but challenging, as it is labor-intensive and thus time-consuming.
This paper investigates the application of large language models (LLMs) in the domain of binary code semantic understanding. 
While these models have achieved groundbreaking results in tasks like natural language processing and source code summarization, their efficacy in decoding binary code remains an unexplored territory. 
We have constructed an extensive binary code summarization dataset including over 557K binary functions in different binary code representations from various binaries.
We also devise a novel method of prompt synthesis and optimization to generate and select the optimal prompts.
To provide a precise measurement of LLM-generated summaries, we design a semantic similarity metrics that overcome the limitations of existing exact matching-based metrics.
Our large-scale study of four state-of-the-art LLMs, including ChatGPT, GPT-4, Llama 2, and Code Llama, uncovers 10 key results (\rnumber{1}-\rnumber{10}). 
The results underscore both the transformative potential of LLMs in this field and the challenges yet to be overcome.
\Xin{further polish}
}

%% file: sections/introduction_new.tex

Binary code comprehension is pivotal for binary analysis and reverse engineering, which delve into the intricate details of software to understand its functionalities~\cite{votipka2020observational}. 
While the scope and outcomes of reverse engineering can differ based on tasks such as malware analysis~\cite{or2019dynamic,or2019dynamic}, vulnerability identification~\cite{gao2018vulseeker,luo2023vulhawk}, and symbol recovery~\cite{jin2022symlm,pei2021stateformer}; the underlying objective remains consistent: to reconstruct the program's logic and garner insights tailored to task-specific goals~\cite{mantovani2022re}. 
Mirroring this intersection of code comprehension and natural language, VirusTotal recently introduced ``Code Insight'', which employs Google's Sec-PaLM to produce natural language summaries of potentially malicious code segments, enhancing threat understanding and detection for security analysts~\cite{virustotal-blog-2023}.  \looseness=-1

\ignore{
Binary code analysis is pivotal in the realm of computer security, holding immense significance as the backbone of our increasingly software-dependent world. Whether it is the critical systems safeguarding our data or the ubiquitous apps that pepper our digital lives, delving into the underlying code is indispensable. This analytical process acts as a critical interface between human creativity and the rigid exactitude of machine language, serving as a linchpin in both offensive and defensive security efforts~\cite{shoshitaishvili2016sok}. In the domain of defensive security, thorough binary code analysis aids in identifying vulnerabilities~\cite{eschweiler2016discovre,luo2023vulhawk}, uncovering hidden backdoors~\cite{shoshitaishvili2015firmalice}, and ensuring that software behavior aligns with its intended design. Conversely, in offensive security, understanding binary code is crucial for exploiting vulnerabilities, developing sophisticated malware~\cite{mohan2012frankenstein}, and advancing penetration testing techniques~\cite{szekeres2013sok}. Yet, to be genuinely transformative in these realms, a deep, semantic understanding of binary code is vital---one that extends beyond mere pattern recognition and delves into the embedded logic, intent, and function. \looseness=-1
}

Although significant progress has been made in automating various binary analysis tasks~\cite{shoshitaishvili2016sok, baldoni2018survey}, reverse engineering of binary code semantics largely remains a process where human expertise is indispensable~\cite{mantovani2022re}.
This task is challenging, demanding considerable skill and time investment, especially given the absence of high-level semantics such as symbol names and data types. 
For example, it took about 40 minutes on average for 31 reverse engineering professionals to understand malicious decompiled code with less than 150 code lines~\cite{yakdan2016helping}.
Efforts such as the DARPA Cyber Grand Challenge (CGC)~\cite{darpa-cyber-grand-challenge} have propelled advancements in computers' capacity to autonomously reason about program binaries. 
However, these automated solutions are still distant from being able to rival the expertise of human reverse engineering experts.

\ignore{
The recent surge in computational capabilities, primarily due to the advent of ``large language models'' (LLMs) like ChatGPT and GPT-4, offers a potential avenue to address these complexities. These innovative models represent significant strides towards realizing the ambitions of strong artificial intelligence, boasting capabilities that enable interactions mirroring human-like intelligence and understanding. The potentialities of LLMs like ChatGPT have already sparked discussions about revolutionary shifts in software engineering and source code analysis based tasks (e.g., Fuzzing~\cite{}, repair~\cite{}), painting a picture of a future where software development and analysis are profoundly shaped by advanced language models.
\Xin{ChatGPT has exhibited promising capacities in software engineering. Cite the papers.}
}

Recent advances in machine learning, particularly the emergence of large language models (LLMs) such as ChatGPT and GPT-4~\cite{openai2023gpt}, offer a promising avenue for addressing these challenges.
These models have unlocked the potential for ``strong artificial intelligence'', allowing interactions that closely mimic human-like intelligence and comprehension.
The capabilities of LLMs have already initiated groundbreaking transformations in binary analysis tasks. 
Notable examples include binary symbol recovery \cite{chen2022augmenting, jin2022symlm} and memory alias analysis~\cite{pei2022neudep}. 
Additionally, LLMs, particularly advanced ones like ChatGPT have demonstrated promising results in text/source code comprehension and summarization~\cite{tian2023chatgpt, yang2023exploring}.
The emergence and widespread adoption of LLMs across various domains pose a pressing question: \textit{Can large language models, including ChatGPT, GPT-4, and other LLMs, effectively summarize the semantics of binary code?}    \looseness=-1

\ignore{
However, the journey to harness the full potential of LLMs in binary code analysis is far from straightforward. These models, developed by OpenAI, are predominantly trained on publicly accessible data, including massive amount of textual content and code repositories. A notable and critical void in their training data is the sparse presence of binary code, raising significant questions about their proficiency in deciphering binary code semantics. This scarcity of binary code in training data potentially caps the models' comprehension abilities in binary code analysis, leaving a vital area in software engineering largely untouched and brimming with unexplored potentials. The unexplored territory between LLMs and binary code analysis necessitates an exploration aimed at revealing the extent to which models like GPT-4 and ChatGPT can navigate the intricacies of binary code.
}

\subsection{Challenges}
\label{sub:challenges}
The task to assess the full potential of LLMs in binary code summarization is far from straightforward due to the following challenges. \looseness=-1


\paragraph{Lack of real-world dataset}  
First, unlike readily accessible textual and source code corpora (\eg, CommonCrawl~\cite{commoncrawl} and BigCode~\cite{kocetkov2022stack}), the absence of a comprehensive and large-scale binary code summarization dataset with ground truth, encompassing various binaries that conform to real-world compilation settings, poses a significant obstacle.
Additionally, while existing LLMs only accept text-like input, binary code can be represented in different formats, such as assembly and decompiled code, making it challenging to ascertain the most suitable input for LLMs. 

\paragraph{Semantic gaps in black-box models}
Models like ChatGPT and GPT-4 undergo pretraining on an extensive corpus of textual data with sparsely represented binary code.
It necessities mitigating the semantic gap between pretraining and our task, \ie, binary code summarization.
Also, LLMs are commonly adapted to downstream tasks via either fine-tuning or prompting strategies~\cite{liu2023pre}.
However, finetuning LLMs are resource-intensive thus computationally infeasible, especially for the closed-source ChatGPT and GPT-4.
For prompting, it has been found that instructing LLMs with
different prompts can yield distinct performance~\cite{zhou2022large}.
For our task, devising effective prompts to maximize LLMs' performance remains challenging considering the black-box nature of LLMs, making it hard to interpret what features or words contribute the most to the summarization results.

\paragraph{Varied outputs for the same semantics}
LLMs produce outputs in natural languages, and thus the generated binary code summarization is in fuzzy textual format with semantically similar but syntactically different expressions, \eg, acronyms and synonyms.
Precisely evaluating such summarization results is necessary but presents its own set of difficulties.
For instance, the same semantics can be expressed by different words, phrases, and sentences in ground truth and LLM-generated summaries. \looseness=-1
    
\ignore{Addressing the aforementioned limitations and uncertainties, this paper proposes a systematic exploration of GPT-4’s and ChatGPT’s abilities to understand binary code. The endeavor is fraught with challenges, owing to the nuances of binary code and the gaps in the training data of current LLMs. The exploration strives not only to bring clarity to the capabilities of these advanced models in binary code analysis but also to spotlight innovative methodologies and approaches that could potentially amplify their effectiveness in this domain. This measurement investigation is geared towards uncovering deeper insights and paving the way for leveraging LLMs in binary code analysis, possibly unlocking new paradigms in software engineering and security.}

\subsection{Insights}
\label{sec:insights}
Addressing the aforementioned challenges, 
we propose a systematic exploration of LLMs' capacities in binary code comprehension. Specifically, our study builds upon the following key insights:

    
\paragraph{Utilizing code comments as ground truth with various binary representations}
Developers typically annotate source code with comments to elucidate both the programmer's intent and the code's functionality, and consequently most existing research on source code summarization~\cite{shi2022evaluation} and binary code summarization~\cite{al2023extending} utilizes code comments as the ground truth.
Moreover, the commercial off-the-shelf (COTS) binaries are compiled
with various settings, \ie, cross computer architectures and optimization levels.
With such diverse binaries, existing binary analysis efforts have used different binary code representations, \eg, raw bytes in XDA~\cite{pei2020xda}, assembly code in SymLM~\cite{jin2022symlm}, and decompiled code in DITRY~\cite{chen2022augmenting}.
Thus, we have constructed a large-scale dataset using code comments as ground truth summaries with binaries compiled into 4 architectures and 4 optimization levels, from which 4 different binary representations are further generated.

\paragraph{Exploiting in-context prompt synthesis and optimization}
For our task, prompts crafted by experienced binary analysis or prompt engineering professionals may yield commendable performance of LLMs.
Nevertheless, this approach is constrained by its subjectivity and may not fully optimize LLM performance.
Meanwhile, LLMs have recently shown the great protential of automatic prompt systhensis and optimization.
Specifically, LLMs have the capacity akin to that of human prompt engineers~\cite{zhou2022large} and they exhibit the ability to self-improve~\cite{huang2022large}.
Therefore, we propose to first search over a pool of prompt candidates that are generated by LLMs themselves in an in-context manner, then optimize the LLM-generated and human-crafted prompts, and finally select the optimal prompts that achieve the best performance evaluated on our task. \looseness=-1

\paragraph{Using semantic-embeddings to calculate the similarity}
The commonly used metrics (\eg, BLEU~\cite{papineni2002bleu}, METEOR~\cite{banerjee2005meteor}, and ROUGE \cite{lin2004rouge}) for evaluation code summarization predominantly focus on exact matching of n-grams.
However, they fail to capture the same semantics in acronyms, synonyms, and syntactically different but semantically similar expressions.
The assessment of LLM-generated binary code summaries should be performed at the semantic level. Therefore, we propose to evaluate LLMs' performance by calculating similarity using semantic embeddings.

\subsection{Contributions and Findings} 
These insights have guided our study design, resulting in 10 key measurement results (\rnumber{1}-\rnumber{10})\add{.} \remove{(and 6 additional findings (\fnumber{1}-\fnumber{6})).} In the following, we outline the key contributions:
\begin{packeditemize}
\item {\bf A large-scale, open source, and comprehensive binary code summarization dataset}: 
We have curated 44 open source projects (with totally 11,475,734 lines of code), which are compiled into binaries across 4 architectures (\ie, x86, x64, ARM, and MIPS) and 4 optimization levels (O0-O3). 
With our comment extraction and function-comment matching, we obtain \totalFunNum binary functions with comment-based ground truth summaries. 
We further generated four binary code representations, including raw bytes, assembly code, intermediate representations (IR), and decompiled code (both with and without debugging symbols), along with their source code. \looseness=-1


\item {\bf Novel techniques}:
To generate optimal prompts that maximize LLM performance, we have devised a four-step procedure, including in-context prompt synthesis, prompt variant generation, prompt optimization, and task-specific prompt selection.
Moreover, we also introduce a novel semantic embedding-based evaluation metric by calculating the semantic similarity between ground truth and LLM-generated summaries for our task by semantic embeddings generated from a pretrained LLM.
\item {\bf Extensive measurement}:
We have focused on four state-of-the-art LLMs: GPT-4, ChatGPT (GPT-3.5), Llama 2~\cite{touvron2023llama}, and Code Llama~\cite{roziere2023code}, that have shown significant impact and advanced performance in downstream tasks along with the existing binary code summarization model BinT5~\cite{al2023extending}.
Our four critical sets of evaluations generate \totalTokens tokens in total with a cost of 11,418 US dollars and 873 NVIDIA A100 GPU hours, resulting in the following findings.
First, stripping debugging symbols from binaries can lead to significant semantic loss (55.0\%) (\rnumber{1}), and decompiled code is the best representation for LLMs to understand binary code (\rnumber{2}).
Second, among the LLMs, ChatGPT and Code Llama perform the best in understanding binary code when with and without debugging symbols, respectively (\rnumber{3}).
Code Llama consistently outperforms Llama 2, the base model from which Code Llama is fine-tuned, but merely fine-tuning CodeT5 on decompiled code does not yield satisfactory performance for BinT5 (\rnumber{4}). 
Additionally, our efficiency study reveals that ChatGPT and Llama 2 models are up to 2.9$\times$ and 1.5$\times$ faster than GPT-4 and Code Llama models, respectively (\rnumber{5}).
Third, among the four computer architectures, LLMs perform the best on the x64 and MIPS binaries (achieving up to 16.0\% better scores) given decompiled code with and without symbols, respectively (\rnumber{6}).
However, there are marginal performance gaps for LLMs at different optimization levels (\rnumber{7}).
Evaluations of the outputs of different decompilers (Ghidra, Hex-Rays, and Angr) demonstrate that LLMs perform the best on Hex-Rays decompiled code with a 60.7\% better score (\rnumber{8}).
We further investigate which symbols (\ie, function names, variable names, and data types) contribute the most to binary code semantics and find function names contribute the most to this semantics (\rnumber{9}).
Finally, we also study the different prompt engineering techniques, \ie, zero-shot, few-shot, and chain-of-thought prompting, for our task, and identify zero-shot prompts as the best jointly considering the performance and cost (\rnumber{10}).
\remove{In addition to the measurement results, we have also obtained six findings (\fnumber{1}-\fnumber{6}) by case studies (Appendix \S\ref{case:study}, \S\ref{sec:gpt-4-result}, and \S\ref{sec:summary-manipulation}), regarding to what extent LLM can understand binary code, GPT-4 summaries, and LLM summarizaition vulnerabilities. Our code and dataset are released at \url{https://github.com/xinjin95/BinSum}.}
\end{packeditemize}

\add{
\paragraph{Data Availability and Supplementary Material}
Upon publication, we will make our artifact and dataset publicly available.
Due to space constraints, additional details, such as our dataset specifics, synthesized prompts, further findings, and extended evaluation information, are included in the online anonymous supplementary material: {\url{https://rb.gy/1idwcc}}.
}

\ignore{ 

\ZQ{I would merge these results into the ones above.}

\noindent {\bf Summary of Findings (\fnumber{1}-\fnumber{4})}:
Our case studies of LLM-generated summaries provide additional confirmation of the deterioration of binary code semantics caused by symbol stripping (\fnumber{1}).
Furthermore, rather than elucidating high-level functionalities, LLMs tend to generate operational summaries for low-level binary representations, e.g., assembly code (\fnumber{2}).
In the case of raw bytes, it is intriguing to note the implicit code-lifting behavior exhibited by LLM-generated summaries (\fnumber{3}).
And our proposed semantic evaluation metric has shown to be better suited for our task, as it effectively captures summary semantics (\fnumber{4}). 
}

\ignore{
Our comprehensive experiments and analyses bring to the fore nuanced insights and a clearer understanding of the capabilities and limitations of LLMs in binary code analysis. The outcomes of this study are multifaceted, highlighting promising aspects and pinpointing areas necessitating further refinement and research. We delve into how models like GPT-4 and ChatGPT, despite their groundbreaking generalization abilities, encounter unique challenges when applied to binary code, offering a detailed exposition of their performance and the possible enhancements needed. The findings from this study serve as a beacon, illuminating the path forward in the convergence of large language models and binary code analysis, and provide a foundation upon which future innovations can build, potentially reshaping the landscape of binary code analysis and software engineering.
\Xin{Sell the contribution of dataset.}

\paragraph{Contributions} This paper makes the following key contributions.

\begin{packeditemize}

    \item \textbf{Unprecedented Exploration and Analytical Depth:}
    We conduct a groundbreaking exploration and in-depth analysis of the potentials of Large Language Models (LLMs), particularly GPT-4 and ChatGPT, in binary code analysis, revealing the complexities and offering insights into this novel and intricate domain.
    
    \item \textbf{Methodological Innovation and Empirical Insights:}
    We pioneer innovative methodologies and conduct rigorous empirical evaluations to understand and enhance the applicability of LLMs in binary code. Our findings provide nuanced insights into the strengths and limitations, suggesting practical enhancements to leverage LLMs effectively in software engineering and computer security.
    
    \item \textbf{Comprehensive Framework and Practical Implications:}
    Our work establishes a comprehensive framework for understanding the interplay between LLMs and binary code analysis, offering detailed implications and serving as a foundational reference for future theoretical and practical endeavors in diverse application areas, including software development and defensive security.
\end{packeditemize} 
}

\ignore{

The recent introduction of ChatGPT and GPT-4, powerful generative language models, has generated significant interest within the binary reverse engineering community. These models have been hailed as a major step towards achieving the long-desired goal of strong artificial intelligence, enabling human-like interactions and showcasing impressive intelligence capabilities.
Initial reports have suggested that OpenAI models, including ChatGPT, have the potential to revolutionize the field of software engineering, signaling a promising future for LLM-driven software engineering.
However, the specific application of GPT-4 and ChatGPT in the realm of binary code understanding remains largely unexplored. These OpenAI models are trained on publicly available data sources, encompassing online textual data and code repositories. Unfortunately, binary code data is rarely present in such sources, which may limit the models' abilities in comprehending binary code.
To address this limitation, we propose an investigation into the binary code understanding capacity of GPT-4 and ChatGPT. By exploring their capabilities in this domain, we aim to shed light on their potential in enhancing binary code analysis and comprehension.

Inspired by the success of LLMs in the fields of source code summarization and natural language processing (NLP), we aim to study the effectiveness of existing LLMs in comprehending the semantics of binary code. 
This area of investigation, to the best of our knowledge, has not yet been explored.
}

%% file: sections/design.tex

In this section, we present the detailed design of how we measure the binary code summarization capability of LLMs.  Note that, similar to existing research~\cite{shi2022evaluation, wang2021codet5, feng2020codebert, al2023extending}, we focus on summarizing binary code at the function level. 
For clarity, we formalize this task as: Given a binary function $f$ extracted from binary executables and represented as a textual representation $x_f\in \mathcal{X}$, which can manifest as raw bytes, assembly code, intermediate representation (IR) code, or decompiled code, we concatenate it with a predetermined prompt $I$, resulting in an augmented input $x_f^\prime=\{I;x_f\}$. Using this input, a binary code summarization system, \( \mathcal{M}: \mathcal{X} \rightarrow \mathcal{Y} \), generates a natural language summary \( y \in \mathcal{Y} \), encapsulating the semantic crux of \( f \).  \looseness=-1

\ignore{
To formalize this task, we define it as follows:

Given a binary function $f$ extracted from binary executables, it is represented as a textual representation $x_f\in \mathcal{X}$. This representation could take the form of raw bytes, assembly code, intermediate representation (IR) code, or decompiled code.
To initiate the summarization process, a predetermined prompt $I$ is concatenated with $x_f$, resulting in an augmented input $x^\prime=\{I;x_f\}$.
Given $x^\prime$, a binary code summarization system, \( \mathcal{M}: \mathcal{X} \rightarrow \mathcal{Y} \), generates a natural language summary \( y \in \mathcal{Y} \), capturing the semantic essence of \( x_f \).
Similar to existing research on source and binary code summarization \cite{shi2022evaluation, wang2021codet5, feng2020codebert, al2023extending}, our focus is on summarizing binary code at the function level. \looseness=-1
}

\ignore{
Note that we focus on summarizing binary code at the function level.
Our design choice is based on two observations. 
First, functions are the basic and reusable building blocks of binary executables that encapsulate specific functionality as a whole.
Second, functions often have well-defined boundaries, and existing binary function boundary detection research has achieved considerable accuracy (over 99\%)~\cite{pei2020xda}.
Based on the similar observations, existing source and binary code summarization research~\cite{shi2022evaluation,wang2021codet5,feng2020codebert,al2023extending} mostly focus on function-level summarization.
}

Our study of the binary code summarization task follows three pivotal steps:
(\textit{i}) We first construct a comprehensive binary code summarization dataset by compiling open-source repositories, resulting in diverse binaries across different architectures and optimization levels.
These binaries are then further processed to yield four distinct binary code representations.
(\textit{ii}) Next, we follow a four-step procedure to synthesize, optimize, and select the most suitable prompts, guided by task-specific evaluations.
(\textit{iii}) Finally, we evaluate four state-of-the-art generative LLMs and the existing binary code summarization model (BinT5 \cite{al2023extending}) on our dataset, using our proposed semantic summary evaluation metric.
In the subsequent subsections, we provide the detailed design for these three steps. \looseness=-1


\ignore{
\paragraph{Scope} In this paper, we focus on the binary code summarization at the function level.
Functions are the building blocks in both source code and binary code, and it's easier to identify and map developer-written code comments from source functions to binary functions compared to the lower- and higher-level code snippets, e.g., code blocks.
Moreover, we target the generative LLMs, including ChatGPT, GPT-4, Llama 2 and Code Llama.
Our design choice is based on the observation that such LLMs have exhibited impressive generalizability and superior performance in both natural language processing and code comprehension tasks, e.g., source code summarization and code generation tasks~\cite{ma2023scope}.
For evaluating binary code summarization, LLMs are given raw-bytes, assembly code, intermediate representation (IR), and decompiled code as the representations of binary functions.
}


\subsection{Dataset Construction}
\label{sec:dataset}


\begin{figure} 
    \centering
    \includegraphics[width=0.47\textwidth]{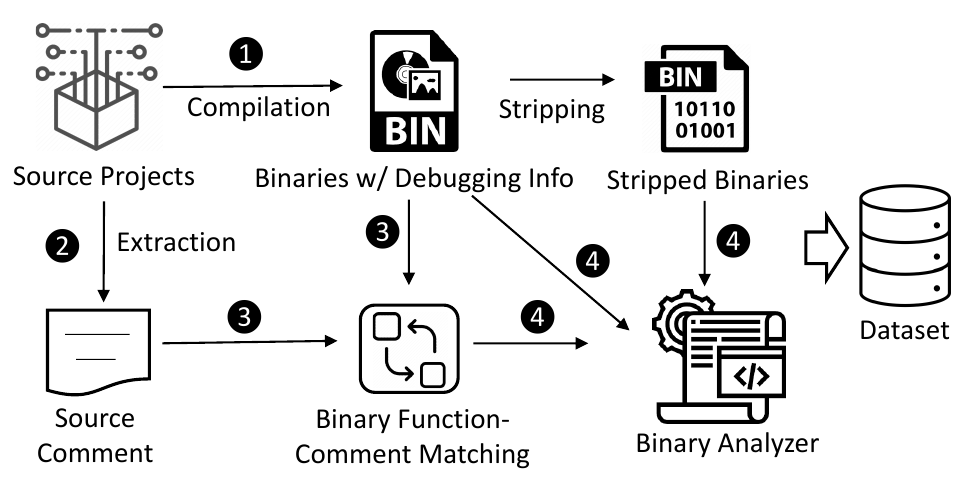}
    \caption{Binary Summarization Dataset Construction}
    \label{fig:dataset-construction}
\end{figure}

Our study aims to gain a comprehensive understanding of LLMs, which requires an extensive binary code dataset with ground truth.
Before constructing such a dataset, we explored existing ones that could be readily employed.
To the best of our knowledge, BinT5 is the only work studying the same task as ours~\cite{al2023extending} but with a focus on decompiled code.
Its released data
\footnote{\url{https://github.com/AISE-TUDelft/Capybara-BinT5}} 
only includes x86 binaries, with an imbalanced distribution in terms of optimization levels, \ie, a rough ratio of 1:2:6:2 for O0-O3 binaries.
We conclude that such a dataset, while useful, falls short of meeting our objectives of obtaining profound insights. 
Consequently, we opt to construct a dataset from scratch to align with our research goals, as shown in \autoref{fig:dataset-construction}. \looseness=-1

\ignore{
Compared to binary code, source code carries far richer semantic information, e.g., variable and function names, thus existing source code models have exhibited promising performance on code summarization.
To obtain code summary references for binary code, an intuitive approach is to generate code summaries using existing source code summarization models and systems, e.g., Github Copilot~\cite{chen2021evaluating}, and then label binary code with generated summaries.
However, such artificial code summaries can contain errors and carry model-specific features~\cite{shi2022evaluation}.

BinT5 is a binary code summarization tool finetuned from \texttt{codet5-base} model. 
However, it only releases the decompiled code dataset without the binaries, which hinges us to obtain the other binary code representations, e.g., assembly code.
Moroever, it only focused on Ghidra-generated decompiled code, so the evaluation on BinT5 dataset can only reveal the Ghidra-specific performance of LLMs, while
we have observed that the code generated from different compilers is dramatically different. 
In addition, BinT5 dataset only covers x86-32 binaries, which are extracted from BinSwarm~\cite{ahmed2021learning}.
Finally, BinT5 dataset is imbalanced regarding optimization levels, i.e., it's binaries with different optimization levels following abnormal distribution.
For example, the decompiled functions of -O0, -O1, -O2, and -O3 optimization levels follow the rough ratio of 1:2:6:2.
Consequently, training on such an imbalanced dataset does not give us a fair performance comparison across different optimization levels.
\Xin{Maybe it's better to revise the above to motivate why we need \textit{large-scale} and \textit{comprehensive} dataset with \textit{ground-truth reference}.}
}

\paragraph{Binary Compilation and Stripping}
Our initial step involves curating public source projects from GNU Software~\cite{GNUSoftware}, including well-known ones that are extensively used by prior works~\cite{jin2022symlm,pei2021stateformer,li2021palmtree}, such as \texttt{coreutils}, \texttt{binutils}, and \texttt{findutils}.
\add{The comprehensive list of 44 GNU C projects includes a total of 11,475,734 lines of code.}
\remove{The comprehensive list of 44 GNU projects is presented in \autoref{tab:source-projects}, encompassing a total of 11,475,734 lines of code.} 
The reason for choosing these GNU software projects is that they have been well-developed and maintained. We then compile them into four architectures (x86, x64, ARM, and MIPS) and four optimization levels (O0-O3) using GCC-7.5, resulting in 8,760 unique binaries with debugging information.
Next, we generate another 8,760 stripped binaries by removing symbols.
\remove{It is worth noting that for ARM and MIPS binaries, we utilize cross-compiler and stripping tools as part of the process (details in \S\ref{sec:cross-compiler}).}





\begin{figure} 
    \centering
    \includegraphics[width=0.5\textwidth]{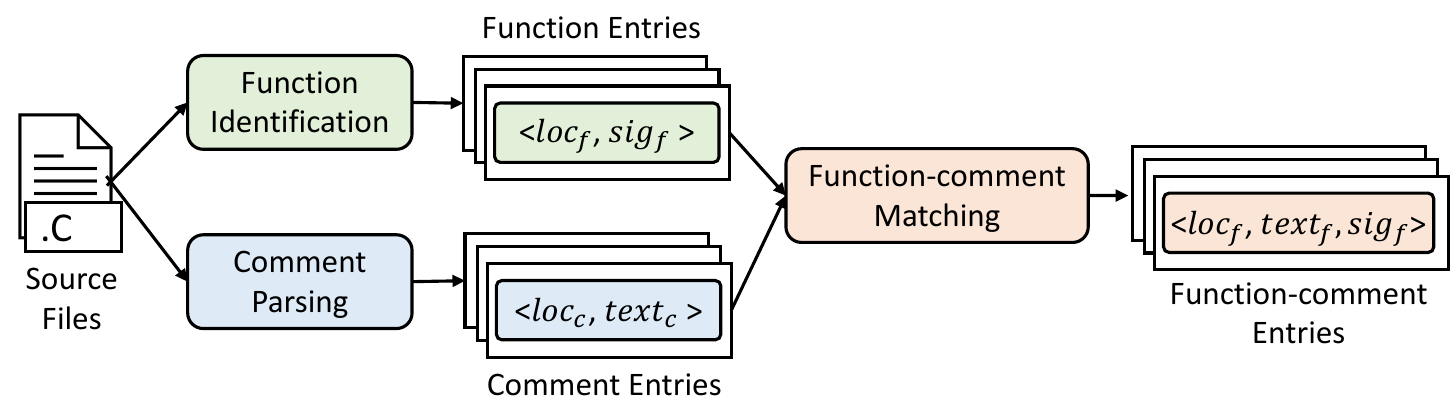}
    \caption{Source Code Comment Parser}
    \label{fig:comment-parser}
\end{figure}

\add{
\paragraph{Source Code Comment Extraction}
As alluded in \S\ref{sec:intro}, we have decided to use developer-written comments as ground truth.
However, our binaries do not contain any comments, as the comments have been removed by the compiler's preprocessor.
Therefore, we need to extract code comments from the source code to annotate binary functions, which is a completely new task, thus no publicly available tools have been designed for this.
For this, we first build an automated source code comment parser, as illustrated in \autoref{fig:comment-parser}, upon srcML~\cite{collard2013srcml} and ANTLR~\cite{parr2013definitive}. Specifically, our manual study of 450 GNU source functions revealed that:
(\textit{i}) Comments in header files usually recur in corresponding source file function definitions;
(\textit{ii}) C function comments are predominantly placed above their signatures, using `//' and `/**/' for single- and multi-line comments.
Given these observations, our parser targets function and comment extraction from source files. The function identification module constructs abstract syntax trees (ASTs) via ANTLR, converts ASTs to XML documents with srcML, and locates XML-tagged functions, outputting function signatures ($sig_f$) and line ranges ($loc_f$).
In parallel, the comment parsing module uses regex to identify comment texts ($text_c$) and their line ranges ($loc_c$). To associate functions with comments, we pair comments and subsequent functions whose line numbers are adjacently sequential, yielding matched function-comment entries. 
\remove{It is worth noting that we exclude internal function comments (\eg, comments at lines 17 and 21) from our analysis. Because such comments do not fully encapsulate the semantics of the function. Additionally, mapping these comments to corresponding blocks in binary code is exceptionally challenging, due to the lack of indicative symbols.}
\looseness=-1
}

\remove{
\paragraph{Source Code Comment Extraction}
As alluded in \S\ref{sec:intro}, we have decided to use developer-written comments as ground truth.
However, our binaries do not contain any comments, as the comments have been removed by the compiler's preprocessor.
Therefore, we need to extract code comments from the source code to annotate binary functions, which is a completely new task, thus no publicly available tools have been designed for this.
As such, we first build a source code parser to achieve this.
Specifically, for each binary function in the DWARF entries of binaries with debugging information, we pinpoint and use the comment of its corresponding function in the source code as the ground truth. \looseness=-1

However, this process requires the precise identification of both comments and source code functions, especially function names, which presents several challenges. First, developers typically compose comments not only for functions but also for other code elements, such as macro definitions, statements, and basic blocks, as shown in~\autoref{fig:comment-example}.
It becomes difficult to discern comments that are specifically intended for functions. For instance, 
only the comment at lines 11 and 12 in  \autoref{fig:comment-example} is the desired comment for the target function \texttt{yylex\_destroy}. Moreover, rather than defining signatures in one line, this example shows a multiple-line signature definition (lines 13 to 15), which means that a simple text-parsing-based method may fail to identify the desired function name.
Furthermore, in addition to commenting in the source (\texttt{.c}) files, developers also create comments in the header (\texttt{.h}) files, which further complex the problem. \looseness=-1

\input{figures/comment-parsing-challenge}

Fortunately, we have developed an automated source code comment parser (step \ding{183} in \autoref{fig:dataset-construction}) to address all these challenges, as illustrated in \autoref{fig:comment-parser}, upon srcML~\cite{collard2013srcml} and ANTLR~\cite{parr2013definitive}. Specifically, a manual study of 450 GNU source functions revealed that:
(\textit{i}) Comments in header files usually recur in corresponding source file function definitions;
(\textit{ii}) C function comments are predominantly placed above their signatures, using `//' and `/**/' for single- and multi-line comments.
Given these observations, our parser targets function and comment extraction from source files. The function identification module constructs abstract syntax trees (ASTs) via ANTLR, converts ASTs to XML documents with srcML, and locates XML-tagged functions, outputting function signatures ($sig_f$) and line ranges ($loc_f$).
In parallel, the comment parsing module uses regex to identify comment texts ($text_c$) and their line ranges ($loc_c$). To associate functions with comments, we pair comments and subsequent functions whose line numbers are adjacently sequential, yielding matched function-comment entries. 
\remove{It is worth noting that we exclude internal function comments (\eg, comments at lines 17 and 21) from our analysis. Because such comments do not fully encapsulate the semantics of the function. Additionally, mapping these comments to corresponding blocks in binary code is exceptionally challenging, due to the lack of indicative symbols.}
\looseness=-1
}

\ignore{
Automatically extracting such code-comment information, however, presents several formidable challenges.
Firstly, developers compose comments not only for functions but also for other code elements, such as macro definitions, statements, and basic blocks, as shown in~\autoref{fig:comment-example}.
It becomes difficult to discern comments that are specifically intended for functions.
Secondly, the idiosyncratic programming conventions of different developers give rise to variations in how functions and comments are defined. 
For instance, while some developers encapsulate the entire function signature within a single line, others opt for a multi-line format, such as the signature of function \texttt{yylex\_destroy} in \autoref{fig:comment-example}.
This diversity in formatting makes it challenging to consistently identify the complete function signature and distinguish function definition and other code elements.
For example, in \autoref{fig:comment-example}, the mixed use of preprocessor directives, such as \texttt{\#define} at line 3 and \texttt{\#ifdef} at line 13, make it difficult to distinguish the function and macro definitions.
However, we found that none of the existing source code parsers can resolve such challenges.
For instance, \texttt{pycparser}, a popular C source code parser, cannot identify code comments as it requires the preprocessed source files where code comments have already been removed, while it also needs dependency library files which are not always available for us. 
}

\ignore{
To tackle these challenges, we initiated a manual examination of 450 randomly selected source functions drawn from diverse source projects. Our investigation revealed the following observations: (\textit{i}) Most functions lacked accompanying comments; 
(\textit{ii}) In addition to directly commenting functions within the source files (typically denoted as \texttt{.c} files), developers also sometimes provided comments for function declarations found in header files (often denoted as \texttt{.h} files);
(\textit{iii}) When developers did comment on function declarations within header files, they often reused these comments or appended additional information when commenting on the function implementations within source files;
(\textit{iv}) In the C programming language, comments manifest in two distinct forms: single-line comments and multi-line comments, where single-line comments are initiated by the symbol `//' and persist until the termination of the line on which they commence, and multi-line comments are enclosed by `/*' to signify their inception and `*/' to mark their conclusion;
(\textit{v}) C function comments are always placed right above function signatures;
(\textit{vi}) some functions are defined multiple times with different code comments in the same source file or across multiple source files, such as the main functions. 
}

\begin{figure*}[]
    \centering
    \includegraphics[width=\textwidth]{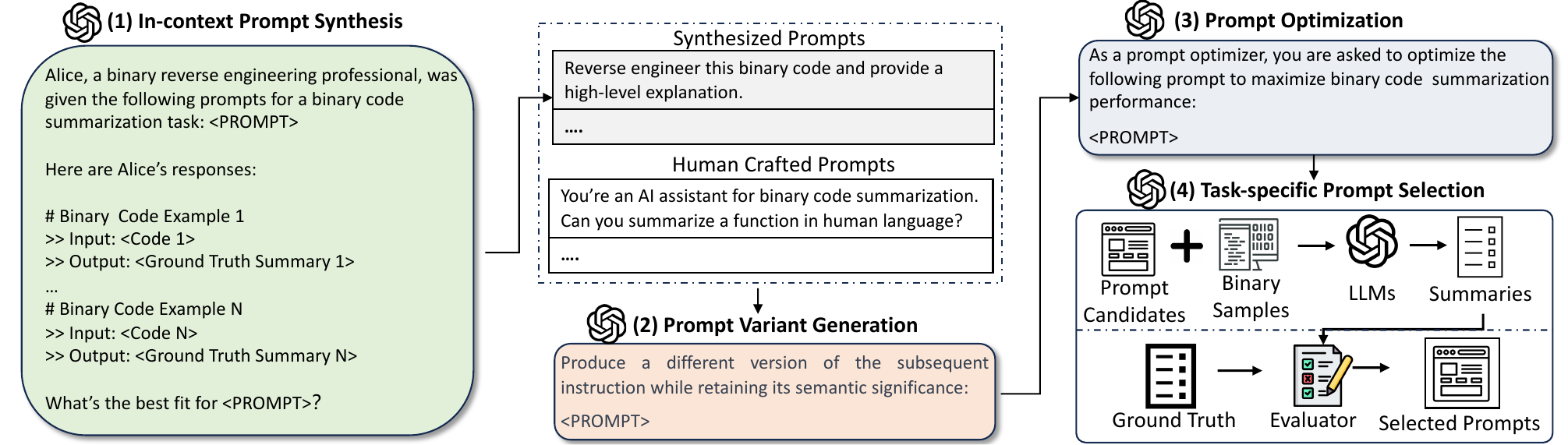}
    \captionof{figure}{Prompt Synthesis and Optimization}
    \label{fig:prompt-synthesis}
\end{figure*}

\ignore{

First, our manual investigation of 450 randomly selected GNU source functions revealed the following observations: 
(\textit{i}) When developers comment on functions in header files, they usually reuse these comments for definitions of the same functions in source files;
(\textit{ii}) C function comments are always placed right above (sometimes with empty lines) function signatures using the symbols `//' and `/**/' for single- and multiple-line comments.
Therefore, our comment parser focuses on parsing functions and comments from source files and generates function and comment entries by its function identification and comment parsing modules in \autoref{fig:comment-parser}.
Specifically, the function identification module first generates abstract syntax trees (ASTs) using ANTLR, then transforms ASTs into XML documents by srmLM, and finally pinpoints the XML-tagged functions.
The results include function entries with function signatures ($sig_f$) and the start/end line numbers of functions ($loc_f$).
The comment parsing module captures comment texts ($text_c$) and their corresponding start/end line numbers ($loc_c$) using regular expressions.
To match the results of these two modules, i.e., functions and comments, we find comments whose end line numbers are exactly smaller than any function's start line number by 1 (empty lines in source files have been removed), resulting in function-comment entries. 
}


\ignore{
For functions that are defined multiple times with different comments, it is not clear which function definition is compiled into binaries.
To avoid ambiguous comments, we remove the functions with ambiguous comments from our dataset.
Overall, our source project parser extracted 40,046 unique functions with comments from the 44 source projects, and 780 functions are with ambiguous comments.
After cleaning, there are 39,266 functions with function-comment entries in our dataset.
}

\paragraph{Function-Comment Matching Between Source and Binaries}
After obtaining source function-comment pairs, we must annotate binary functions with these comments. Our proposed matching approach is based on two insights:
(i) Binaries with debugging information provide function names and addresses via DWARF entries. Given control over the source code, we ensure DWARF symbols are generated and we retain function addresses from non-stripped binaries;
(ii) Internal binary functions possess unique, non-conflicting names, acting as identifiers and facilitating matching with source function-comment pairs.
Based on these insights, Ghidra~\cite{ghidra} is employed to parse DWARF entries, obtaining binary function names and boundaries (start and end addresses). 
We then extract the function name from every source function-comment entry's signature ($sig_f$). 
The final matching process (\ding{184} in \autoref{fig:dataset-construction}) pairs binary function addresses with their related source comments using function names. 
This method identifies comments for 557,664 binary functions across 8,760 binaries with debugging symbols. 
At step \ding{185} in \autoref{fig:dataset-construction}, four representations are generated for binary functions: raw bytes, assembly code, IR code, and decompiled code (detailed in \S\ref{sec:implementation}).
\looseness=-1

\ignore{
\paragraph{Function-Comment Matching Between Source and Binaries}
After obtaining the source function-comment entries, we still need to annotate binary functions with source comments.
For this, we propose a simple but effective matching method based on two observations.
First, for binaries with debugging information, the DWARF entries contain the names and addresses of binary functions. Since we control the source code of the binaries, we can always generate the binaries with DWARF symbols. When stripped the symbols, we also have the matched addresses of the functions from the non-stripped binaries.
Second, the internal binary functions have their own unique names without any conflict.
Therefore, the function names can serve as the identity of binary functions and a bridge to match our identified source function-comment entities.
In particular, for each source project, we use Ghidra to parse DWARF entries of compiled binaries to identify all the binary function names and function boundaries, i.e., start and end addresses.
Next, we iterate every source function-comment entry and extract the function name from its signature ($sig_f$).
Our final step involves matching the function names of the source function-comment entries with the binary function names extracted from the DWARF entries. 
This matching procedure allows us to associate binary function addresses with their corresponding source code comments.
Eventually, our matching process identifies source code comments for 557,664 binary functions from 8,760 binaries with debugging info.
For binary functions, we generate four representations, including raw bytes, assembly code, IR code, and decompiled code using three decompilers (details in \S\ref{sec:setup}).
}









\subsection{Prompt Synthesis and Optimization}
\label{sec:prompt-engineering}

As discussed in \S\ref{sec:insights}, we have observed the great potential of using prompts to guide LLMs to generate summaries in a zero-shot manner.
\ignore{
Driven by the strides made in scaling, LLMs have demonstrated the promising capability to undertake a diverse array of natural language processing (NLP) tasks in a zero-shot manner—meaning they perform these tasks without the need for specific adaptation to downstream data. 
Rather than relying on intricate parameter fine-tuning, a thoughtfully constructed natural language prompt, acting as an instructional directive for an LLM, can be profoundly influential~\cite{liu2023pre}. These prompts essentially steer the LLM, adjusting its behavior, amplifying its capabilities, or honing its performance for specialized task.
Meanwhile, it has been found that LLMs are highly sensitive to the format of the input prompts \cite{wei2023larger}. Notably, prompts that are semantically similar can yield drastically different performance outcomes \cite{kojima2022large, zhou2022large}. 
For instance, the mere inclusion of a phrase like ``Let's think step by step'' has been demonstrated to significantly enhance the performance of InstructGPT in mathematical reasoning tasks \cite{kojima2022large}.
}
However, the ideal prompts for binary code summarization remain implicit but necessary because LLMs are highly sensitive to input prompts~\cite{zhou2022large}.
An intuitive solution is to seek prompts from experienced prompt engineers or binary analysis professionals, but this method has its limitations.
First, human prompt creators might be limited by their knowledge and creativity, leading to suboptimal prompts.
Furthermore, LLMs can exhibit behavior that deviates from human expectations \cite{zhou2022large}.
Lastly, it is not clear whether such prompts can optimize LLMs' performance as LLMs are black-box.
In \autoref{fig:prompt-synthesis}, we propose a four-step solution, encompassing: (1) in-context prompt synthesis, (2) prompt variant generation, (3) prompt optimization, and (4) task-specific prompt selection.
\ignore{An intuitive way to obtain the ideal prompts would be to seek inputs from prompt engineering experts or reverse engineering professionals who can leverage their experience to generate prompts. However, this approach is not without its drawbacks. First, there is uncertainty regarding whether a manually generated prompt can fully optimize the performance of LLMs, primarily because LLMs operate as black-box systems with intricate internal mechanics. 
Second, while employing a variety of prompts can sometimes yield unexpectedly strong LLM performance \cite{jin2021good}, human prompt engineers may be constrained by their existing knowledge and creativity, potentially leading to prompts that do not fully exploit the capabilities of the LLM.
Lastly, it has been noted that LLMs can exhibit behavior that deviates from human expectations \cite{zhou2022large}, making it challenging to predict their responses and evaluate input prompts accurately.}



\paragraph{(1) In-context Prompt Synthesis}
In this approach, we harness LLMs to generate a pool of candidate prompts by themselves within the context of binary code summarization. 
This design choice is inspired by the way human users engage with LLMs when confronted with a new task.
In this scenario, human users often need to experiment with various prompts to elicit the desired responses. 
Due to the virtually unlimited prompt search space, manually generating numerous prompts is very challenging.
Meanwhile, LLMs themselves have shown very promising text generation capacities~\cite{openai2023gpt}, which motivates us to propose an automated prompt synthesis approach. \looseness=-1

Formally, we consider our binary code summarization task specified by the dataset $\mathcal{D} = \{x_i, y_i\}_{i=1}^{N}$ with a prompt targeting model $\mathcal{M}$, where $x_i$ and $y_i$ are the input binary code and the code summary.
Our objective is to generate a suitable prompt, denoted as $p$, in a manner that when $\mathcal{M}$ is given the concatenated input $(p;x)$, $\mathcal{M}$ generates the corresponding output $y$.
We cast this task as an optimization problem to find the optimal prompt $\hat{p}$. 
This optimization is centered around maximizing the expected score $S_{\mathcal{M}}(p, x, y)$, across all potential pairs of $(x, y)$: \looseness=-1

\begin{equation}
    \hat{p} = \operatorname*{arg\max}_{p} S_{\mathcal{M}}(p) = \operatorname*{arg\max}_p\mathbb{E}_{(x,y)} [S_{\mathcal{M}}(p, x, y)]
\end{equation}

\noindent where $S_{\mathcal{M}}(\cdot)$ could be computed in several ways, such as the aggregated token probability~\cite{zhou2022large} and the beam search probability~\cite{li2023guiding}.
However, we observe that API-based LLMs like ChatGPT are black-box generators.
That is, the LLM output does not contain any probability products; thus, these probability-based score functions cannot apply to our task.
Alternatively, we directly take the top-1 LLM-generated prompt as $\hat{p}$ as it is produced by maximizing the autoregressive prediction probability~\cite{openai2023gpt}, which can represent the maximum score of $S_{\mathcal{M}}(p, x, y)$.
\ignore{
Alternatively, we find the top-1\footnote{Top-1 inference results with maximum probability, can be set by model/API parameters, such as \texttt{top\_p=1} and \texttt{n=1} for ChatGPT and GPT-4 chat completion API~\cite{openai-api-chat-complete}.} inference result can represent the maximum score of $S_{\mathcal{M}}(p, x, y)$ because it is produced by maximizing the autoregressive prediction probability~\cite{openai2023gpt}.
Therefore, we directly take the top-1 LLM-generated prompt as $\hat{p}$.
}
Apart from synthesized prompts, we also include human-crafted prompts to increase the size and diversity of our prompt candidate pool as demonstrated in \autoref{fig:prompt-synthesis}. \looseness=-1

\paragraph{(2) Prompt Variant Generation}
In our initial prompt candidate pool, there may be instances where this synthesis approach falls short, either due to a lack of diversity or an absence of candidates with sufficiently high scores.
In response to such problems, we generate prompt variants that are semantically similar to the candidate prompts as shown in \autoref{fig:prompt-synthesis}.
This approach essentially explores the search space in close proximity to the current best candidates. 
Therefore, it enables us to create new prompts that are more likely to yield favorable results. \looseness=-1


\paragraph{(3) Prompt Optimization}
To further enhance the performance of LLMs in the context of binary code summarization, we introduce a strategy to optimize the previously synthesized prompts and their variants using the LLMs themselves.
Our design choice is motivated by prior research, showing that LLMs can optimize the problems of linear regression and traveling salesman \cite{yang2023large}, and LLMs can also self-improve~\cite{huang2022large}.
In our task, the advantage of employing LLMs for prompt optimization lies in their ability to comprehend natural language, \ie, enabling us to convey our prompt optimization objectives by high-level textual instructions without formal specifications. 
Given the candidate prompts, we instruct LLMs to serve as a prompt optimizer by a meta-instruction.
This meta-instruction encapsulates the objective function, \ie, maximizing performance, and solution constraints, \ie, the context of the binary code summarization task, as shown in \autoref{fig:prompt-synthesis}. \looseness=-1

\remove{
\paragraph{(4) Task-Specific Prompt Selection}
After generating a pool of prompt candidates by prompt synthesis and optimization, the subsequent selection step becomes crucial in determining the most promising candidates for our task.
This process closely aligns with the well-established problem of best arm identification in Upper Confidence Bound (UCB) optimization~\cite{pryzant2023automatic}. 
In this context, arms correspond to the prompt candidates, the hidden value of each arm represents its performance on our task, and the arm-pulling action corresponds to evaluating prompts with randomly selected data points. 
The ultimate objective is to identify the top-performing arms while minimizing the number of pull efforts required to achieve this outcome. 
\looseness=-1
}

\add{
\paragraph{(4) Task-Specific Prompt Selection}
After generating a pool of prompt candidates,
the subsequent selection step becomes crucial in determining the most promising candidates for our task.}
Evaluating each candidate prompt on our entire dataset is a resource-intensive process and prohibitively expensive. 
Inspired by previous work that efficiently estimates prompt performance~\cite{zhou2022large}, we assess prompt candidates by evaluating them on a randomly selected subset of data samples from our dataset.
\autoref{fig:prompt-synthesis} presents how we proceed: We first generate the binary code summary by querying LLMs with concatenated every prompt candidate and the test binary sample.
These generated summaries are then evaluated by calculating semantic similarity scores to ground truth (see the evaluator details in \S\ref{sec:testing-eval}). 
Finally, the prompts with the best scores are selected as the final prompts for our subsequent evaluations.

\subsection{Testing and Semantic Evaluation}
\label{sec:testing-eval}

Our goal is to assess existing LLMs' performance on our task. 
For this, we follow a two-step process: (1) LLM testing on binary code summarization and (2) semantic summary similarity evaluation.

\paragraph{(1) LLM Testing on Binary Code Summarization}
\ignore{To assess the binary code understanding capabilities of LLMs, we create test samples by combining selected prompts with the binary code of individual functions. 
Moreover, we generate task-specific test samples to evaluate LLMs on distinct evaluation tasks. 
For instance, when evaluating binary code representations, we generate separate test samples for each binary function, encompassing raw bytes, assembly code, Intermediate Representation (IR) code, decompiled code from stripped binaries, decompiled code from binaries with debugging information, and source code.
We have chosen the most advanced LLMs, including ChatGPT and GPT-4 from the OpenAI GPT model family, as well as the Meta Llama model, which has exhibited dominant performance across numerous NLP and code-related tasks \cite{huggingface-llm-leaderboard, touvron2023llama, openai2023gpt}. 
Specifically, we have opted for the latest version of Llama, namely Llama 2. 
Additionally, we include the recent Code Llama model, which has been fine-tuned on code datasets \cite{roziere2023code}.
}
To generate summaries, we concatenate the selected prompt with the test binary code sample for model input preparation. 
This configuration essentially places us in a zero-shot setting, jointly considering both performance and cost considerations (see \S\ref{sec:eval-RQ4} for the evaluations of different prompting settings).
As LLMs are generative models, they can generate summaries of varying lengths. Nevertheless, excessively verbose and lengthy summaries are generally regarded as low-quality \cite{iyer2016summarizing}. 
Therefore, we impose limits on response length for our target models.
For instance, we explicitly specify these limits in our prompts with the instruction ``summarize ... in $N$ words'', where $N$ is the average number of words in the ground-truth summaries from our dataset. 
We observe that our test models exhibit an adaptive length behavior when generating summaries. 
That is, they tend to produce summaries with lengths that closely hover near the predetermined length limit, demonstrating a capacity to adjust the length intelligently without rigidly and dogmatically adhering to the predetermined length limit.
This is beneficial considering the inherent diversity of binary code samples.
\remove{
\begin{figure} 
    \centering
    \includegraphics[width=0.35\textwidth]{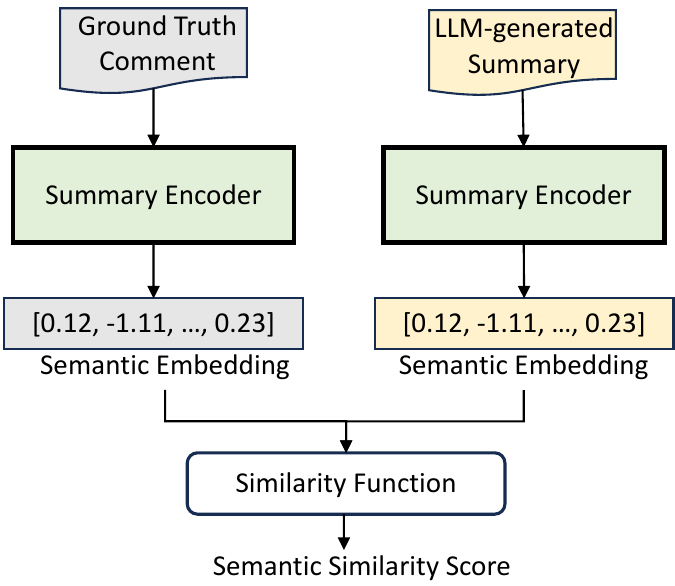}
    \caption{Semantic Summary Similarity Calculation
    }
    \label{fig:semantic-metric}
\end{figure}
}

\paragraph{(2) Semantic Summary Similarity Evaluation}
To assess LLM-generated binary code summaries, we aim to estimate how they align with the ground-truth summaries.
For this, we notice that prior source and binary code summarization research~\cite{shi2022evaluation,al2023extending} both use exact matching-based metrics, \eg, BLEU~\cite{papineni2002bleu}, METEOR~\cite{banerjee2005meteor}, and ROUGE~\cite{lin2004rouge}.
\add{We have found that these metrics fall short in capturing the essential semantics of summaries.}
\remove{We have found that these metrics fall short in capturing the essential semantics of summaries (see \fnumber{4} and Appendix \S\ref{case:study}).}
Alternatively, we propose solving this problem based on semantic measurement.
For this, we observe that pre-trained LLMs, such as BERT and XLNet, have gained success in text semantic modeling tasks~\cite{min2021recent}.
Therefore, we can calculate the semantic similarity between LLM-generated and ground-truth summaries using these pre-trained models as the semantic measurement. \looseness=-1

Formally, given two sequences of words representing a generated summary $S=\{w_1, w_2, ..., w_n\}$ and a ground truth summary reference $\hat{S}=\{\hat{w_1}, \hat{w_2}, ..., \hat{w_m}\}$, the summary encoder $\mathcal{E}$, which is a pretrained LLM, generates semantic embeddings for each token of $S$ and $\hat{S}$, respectively: $\mathcal{E}(S)=\{e_{w_1}, e_{w_2}, ..., e_{w_n}\}$ and $\mathcal{E}(\hat{S})=\{e_{\hat{w_1}}, e_{\hat{w_2}}, ..., e_{\hat{w_m}}\}$, where $e_w\in \mathbb{R}^{N}$ is the token embedding for token $w$.
Then we compute the embeddings $e_S\in \mathbb{R}^N$ and $e_{\hat{S}}\in \mathbb{R}^N$ representing the entire semantics of $S$ and $\hat{S}$ by aggregating their token embeddings:
\begin{equation} \label{eqn:summary-embedding}
    e_{S} = \frac{1}{|S|}\sum_{i=1}^{n} e_{w_i},\quad e_{\hat{S}} = \frac{1}{|\hat{S}|}\sum_{j=1}^{m} e_{\hat{w_j}}
\end{equation}
Here, we use the mean pooling as the aggregation function which has shown superior results in text embedding tasks~\cite{reimers2019sentence}.
Finally, we calculate the semantic similarity between generated summaries and ground-truth summary references by the cosine similarity function following the best practice~\cite{minaee2021deep}:
\begin{equation}    \label{eqn:cosine-similarity}
    sim(S, \hat{S}) = \frac{e_S\cdot e_{\hat{S}}^\top}{||e_S||\times ||e_{\hat{S}}||} = \frac{\sum_{i=1}^{N}e_S^{(i)}\times e_{\hat{S}}^{(i)}}{\sqrt{\sum_{i=1}^N{e_S^{(i)}}^2}\times\sqrt{\sum_{i=1}^N{e_{\hat{S}}^{(i)}}^2}}
\end{equation}

\noindent The semantic similarity score $sim(S, \hat{S})$ quantifies the alignment between generated summaries and summary references. A higher similarity score indicates a better quality of the generated summary.
\remove{\autoref{fig:semantic-metric} illustrates the process of calculating this semantic similarity score.}
\looseness=-1

%% file: figures/comment-parsing-challenge.tex
\begin{figure}[t]
\centering
\begin{minted}[mathescape,
               linenos,
               numbersep=5pt,
               %gobble=2,
               % frame=lines,
               frame=single,
               fontsize=\scriptsize,
               %fontsize=\footnotesize,
               %framesep=2mm,
               xleftmargin=10pt
               ]{c}
/* Return all but the first "n" matched 
  characters back to the input stream. */
#define yyless(n) \
    do \
    { \
        int yyless_macro_arg = (n); \
        YY_DO_BEFORE_ACTION; \
    } \
    while ( 0 )

/* yylex_destroy is for both reentrant 
    and non-reentrant scanners. */
static int 
yylex_destroy  (const char * yybytes, 
                int  _yybytes_len )
{
    /* Pop the buffer stack, 
        destroying each element. */
    while(YY_CURRENT_BUFFER){
        // Delete current buffer
        yy_delete_buffer( YY_CURRENT_BUFFER  );
        YY_CURRENT_BUFFER_LVALUE = NULL;
        yypop_buffer_state();
    }
    ...//skip for simplicity
    return 0;
}
\end{minted}
\caption{Motivating Example of Source Code Comment Extraction Challenges. This figure presents the \texttt{yylex\_destroy} function from Binutils-2.39 with a signature defined across multiple lines (lines 13 to 16). A simple text parsing-based solution will fail to identify the correct and complete function signature and comments.
}
\label{fig:comment-example}
\end{figure}

%% file: sections/implementation.tex
\noindent\textbf{Dataset Construction.}
\remove{As depicted in \S\ref{sec:dataset}, we have built a binary code summarization dataset by compiling 44 open-source projects with 11,475,734 total lines of code (\autoref{tab:source-projects} in Appendix has more details).
Our dataset contains \totalFunNum binary functions with ground truth summaries.
These functions are from various binaries, and their distributions across architectures (x64, x86, ARM, MIPS) and optimization levels (O0-O3) are presented in \autoref{tab:labeled-binary-functions} in the appendix for readers of interest.}
\add{As depicted in \S\ref{sec:dataset}, we have built a binary code summarization dataset by compiling 44 open-source projects with 11,475,734 total lines of code.
Our dataset contains \totalFunNum binary functions with ground truth summaries.
These functions are from various binaries, and their distributions across architectures (x64, x86, ARM, MIPS) and optimization levels (O0-O3).}
For each function in our dataset, we generated four binary code representations along with source code: 
\begin{packeditemize}
    \item \textbf{Raw bytes} are directly extracted from binaries based on the function start and end addresses (identified in \S\ref{sec:dataset}).
    \item \textbf{Assembly code} is generated by parsing ELF binaries using \texttt{capstone}~\cite{capstone-engine} and \texttt{pyelftools}~\cite{pyelftools} libraries. 
    \item \textbf{IR code} is the IDA microcode generated by our IDA plugin built upon the \texttt{ida\_hexrays.gen\_microcode} API~\cite{ida-hexrays-doc} where we have followed prior binary analysis research~\cite{yu2020order,yu2020codecmr} to use the widely adopted IDA microcode as IR code.
    \item \textbf{Decompiled code} encompasses outputs generated by Ghidra, Hex-rays, and Angr, using the \texttt{getDecompiledFunction}~\cite{ghidra-decompiler-doc}, \texttt{idaapi.decompile}~\cite{ida-idaapi-doc}, and \texttt{angr.analyses.Decompiler} \cite{angr-decompiler-doc} APIs, respectively. We have chosen Ghidra and Hex-rays as they are prominent decompilers.
    Angr, on the other hand, has been selected due to its rapid growth and comprehensive capabilities in binary analysis. 
    \item \textbf{Source code} is generated by parsing source files according to functions' start and end line numbers. Note that we exclude the function comments to avoid information leakage to LLMs when testing their capability of summarization.
\end{packeditemize}

\input{tables/model-statistics}

\paragraph{Prompt Synthesis and Evaluation}
\remove{We have generated a set of 320 prompt candidates based on our prompt synthesis and optimization approach.
As mentioned in the task-specific prompt selection section of \S\ref{sec:prompt-engineering}, it is prohibitively expensive and thus impractical to evaluate all prompt candidates on the entire binary summarization dataset.
Hence, we randomly select 1000 binary function samples from our dataset to conduct binary code summarization using each prompt candidate.
\autoref{tab:top-prompts} presents the top 40 prompts ranked by semantic similarity.
We have selected the best prompt for our subsequent evaluations.
For a fair comparison, we employ the same prompt for all models, except for BinT5, which only takes the decompiled code as input (see \S\ref{sec:prompt-evaluation} for more details). \looseness=-1
}
\add{
We have generated a set of 320 prompt candidates based on our prompt synthesis and optimization approach.
As mentioned in the task-specific prompt selection section of \S\ref{sec:prompt-engineering}, it is prohibitively expensive and thus impractical to evaluate all prompt candidates on our entire dataset.
Hence, we randomly select 1000 binary function samples from our dataset to conduct binary code summarization using each prompt candidate.
We then use the best prompt for our subsequent evaluations.
For a fair comparison, we employ the same prompt for all models, except for BinT5, which only takes the decompiled code as input. \looseness=-1
}

\paragraph{Evaluation Metrics}
Our evaluation metrics include the semantic similarity metric that we proposed in \S\ref{sec:testing-eval}.
Specifically, we implemented our evaluation framework on top of Pytorch~\cite{paszke2019pytorch}, Sentence-Transformers~\cite{reimers2019sentence}, Scipy~\cite{virtanen2020scipy}, and Sentencepiece~\cite{kudo2018sentencepiece}.
For the summary encoder model, we use the pre-trained \texttt{all-mpnet-base-v2} model, which performs the best on 14 text embedding tasks~\cite{sbert-pretrained-models}.
To efficiently compute $e_S$ and $e_{\hat{S}}$ in \autoref{eqn:summary-embedding}, we store token embeddings in a Chroma vector database~\cite{trychroma}.
The calculated semantic similarity scores range from -1 to 1 where a score close to 1 indicates a high quality of LLM-generated code summaries.
In addition to our proposed metric, we also calculate and report the scores of BLEU~\cite{papineni2002bleu}, METEOR~\cite{banerjee2005meteor}, and ROUGE-L~\cite{lin2004rouge} to facilitate future work, as these metrics have previously been employed in both source and binary code summarization research~\cite{al2023extending,wang2021codet5,shi2022evaluation}.
The calculation details of BLEU, METEOR, and ROUGE-L are in Appendix \S\ref{sec:exact-matching-metrics}. 
BLEU, METEOR, and ROUGE-L scores are scaled between 0 and 1, where higher values signify superior summarization quality. \looseness=-1
\ignore{
The similarity scores range from -1 to 1. A score close to 1 indicates the generated summaries closely align in semantics with the ground truth, indicating a high quality of binary code summarization. Conversely, a score nearing -1 indicates a stark dissimilarity in semantics between the generated summaries and the ground truth, signifying an opposing meaning or context.

First, they have been used by the existing work, BinT5~\cite{al2023extending}.
Second, considering their wide adoption in code summarization tasks~\cite{wang2021codet5,feng2020codebert,shi2022evaluation} that are similar to our task, we report these scores to facilitate future work.
For readers' interest, we put the calculation details of BLEU, METEOR, and ROUGE-L in appendix \S\ref{sec:exact-matching-metrics}. 
}

\paragraph{Targeted LLMs}
The LLMs of our interest are GPT-4, ChatGPT, Llama 2 (7B), Llama 2 (13B), Code Llama (7B), Code Llama (13B), and BinT5~\cite{al2023extending}, as presented in \autoref{tab:model-stats}, which has exhibited dominant performance across numerous NLP and code-related tasks \cite{huggingface-llm-leaderboard, touvron2023llama, openai2023gpt}.
To test GPT-4 and ChatGPT, we use OpenAI's chat completion API (\texttt{openai.Completion.create})~\cite{openai-api-chat-complete}, which have called \texttt{gpt-4-0613} and \texttt{gpt-3.5-turbo-16k-0613} backend models, corresponding to the 0613 snapshots, released on June 13th, 2023.
To get consistent results, we have set the \texttt{temperature} parameter~\cite{openai-api-chat-complete}, controlling models' diversity and creative output, to 0.1. 
Furthermore, we set both \texttt{top\_p} and \texttt{n} at 1 to obtain the top-1 results~\cite{openai-api-chat-complete}.
Regarding the parameters \texttt{stop} and \texttt{max\_tokens}, we opt for the default values~\cite{openai-api-chat-complete}, allowing the models to autonomously determine when to conclude their responses.
To enhance inference speed, we have implemented multi-threading, utilizing five and six parallel threads for GPT-4 and ChatGPT, respectively, while ensuring compliance with the rate limits set by OpenAI \cite{openai-rate-limits}.

We have donwloaded Llama models from Huggingface, including meta-llama/Llama-2-7b-chat-hf, meta-llama/Llama-2-13b-chat-hf, codellama/CodeLlama-7b-Instruct-hf, and codellama/CodeLlama-13b-Instruct-hf~\cite{meta-llama,code-llama}.
To run these models, we implemented an inference framework based on Pytorch~\cite{paszke2019pytorch}, transformers~\cite{wolf2019huggingface}, Scikit-learn~\cite{pedregosa2011scikit} and DeepSpeed~\cite{rasley2020deepspeed}.
For efficient inference, we enable mixed precision in BF16 and batch size at 6.
We directly use their tokenizers and pretrained model weights.
We set \texttt{temperature} as 0.1 and \texttt{top\_k} and \texttt{top\_p} as 1 to get the best summaries. 
For BinT5, it only accepts decompiled code, thus we only use it for summarizing decompiled code.
We set its inference batch size as 64, using its maximum context size, \ie, 512.


\ignore{
\paragraph{Binary Code Summarization Dataset Construction}


\paragraph{Binary Code Representaiton Generation}
To examine how LLMs can understand the binary code in different representations, we leverage binary analysis toolkits to generate different representations along with the source function code:

\begin{packeditemize}
    \item \textbf{Raw bytes} are directly extracted from binaries ($b_f$) based on the binary function start and end addresses ($addr_f$).
    \item \textbf{Assembly code} is generated by the \texttt{capstone} disassembly framework in the Intel syntax for each binary function. Specifically, we disassembly the functions by first parsing the ELF binary using the \texttt{pyelftools} package and then generate assembly instructions by the \texttt{capstone.Cs.disasm\_lite} API.
    \item \textbf{IR code} is the microcode produced by IDA, which has also been used for other binary reverse engineering tasks~\cite{yu2020order,yu2020codecmr}. For this, we implement an IDA plugin. Given the start and end addresses of target functions, this plugin first generates microcode by the \texttt{ida\_hexrays.gen\_microcode} API and then remove the control characters that are in the range of 0x00-0x1F in ASCII table. \ZQ{Why to choose IDA IR, why not LLVM IR, etc., has to be justified. Maybe you can say IDA IR is the industry state of the art. Also, there is the Ghidra IR, why not to use that? Similarly, why not VEX IR used by angr? So many questions readers will have in mind. We have to somehow justify}
    \item \textbf{Decompiled code} includes the code generated by Ghidra (the popular open-source binary analyzer), Hex-rays (the popular commercial decompiler), and Angr (the popular and fast-growing binary analysis toolkits). For Ghidra, we developed a Ghidra plugin based upon \texttt{ghidra.app.decompiler.DecompInterface} API, in which the binary functions are decompiled with \texttt{getDecompiledFunction().getC()}. For Hex-rays, we build the IDA plugin and decompile functions with the \texttt{idaapi.decompile} API. 
    For Angr, our decompliation scripts first perform CFGFast analysis for static control-flow and function recovery, and then it generates decompiled code with the \texttt{angr.analyses.Decompiler} API. 
    \item \textbf{Source code} is generated by parsing the source files according to functions' start and end line numbers. Note that we exclude the function comments to avoid information leakage to LLMs.
\end{packeditemize}
}


%% file: tables/model-statistics.tex
\begin{table*}[th]
\caption{Models and Execution Statistics. 
}
\resizebox{\textwidth}{!}{
\begin{tabular}{lccccccccc}
\toprule
                                 &                                    &                                          &                                     &                                          &                                   & \multicolumn{3}{c}{\textbf{\# of Tokens}}             &                                 \\\cline{7-9}
\multirow{-2}{*}{\textbf{Model}} & \multirow{-2}{*}{\textbf{Creator}} & \multirow{-2}{*}{\textbf{\# Parameters}} & \multirow{-2}{*}{\textbf{Modality}} & \multirow{-2}{*}{\textbf{\# Max Tokens}} & \multirow{-2}{*}{\textbf{Access}} & \textbf{Input} & \textbf{Output} & \textbf{Total} & \multirow{-2}{*}{\textbf{Cost}} \\
\midrule
GPT-4                            & OpenAI                             & 1.7T~\cite{GPT4migh91:online}                                 & text                                & 32K                                      & API                       & 309,401,126      & 19,678,448        & 329,079,574    & \$10,462.74                    \\
ChatGPT                          & OpenAI                             & 175B~\cite{sciencefocus-chatgpt-parameters}          & text                                & 16K                                      & API                       & 296,524,698      & 16,338,012        & 312,862,710    & \$954.93                       \\
Llama 2 (7B)                     & Meta AI                            & 7B                                       & text                                & 4,096                                    & Open                              & 736,413,994                          & 33,195,771                            & 769,609,765                        & 163 GPU Hours                    \\
Llama 2 (13B)                    & Meta AI                            & 13B                                      & text                                & 4,096                                    & Open                              & 736,413,994                          & 29,699,869                            & 766,113,863                        & 175 GPU Hours                    \\
Code Llama (7B)                  & Meta AI                            & 7B                                       & text and code                       & 100K                                     & Open                              & 736,413,994                          & 40,016,196                            & 776,430,190                        & 253 GPU Hours                    \\
Code Llama (13B)                 & Meta AI                            & 13B                                      & text and code                       & 100K                                     & Open                              & 736,413,994                          & 33,323,576                            & 769,737,570                        & 266 GPU Hours                    \\
BinT5~\cite{al2023extending}                        & Al-Kaswan et al.                   & 220M                                     & code                                & 512                                      & Open                              & 326,007,985                          & 8,456,320                             & 334,464,305                        & 16 GPU Hours
\\
\bottomrule
\end{tabular}
}

\label{tab:model-stats}
\end{table*}

%% file: sections/evaluation.tex
In this section, we present our evaluation results. While there are many research questions (RQ) centered around this measurement, we seek to answer the following ones: 

\begin{packeditemize}
\item \textbf{RQ1}: To what extent can LLMs comprehend binary code?
What input of binary code impacts LLM's output more?
\item \textbf{RQ2}: Which LLM performs the best on binary code comprehension? Which LLM is more efficient than others?
\item \textbf{RQ3}: How do the different computer architectures and optimization levels affect LLMs' performance?
\item \textbf{RQ4}: 
What are additional factors of binary code input influencing LLMs' comprehension capabilities?
\end{packeditemize}

\subsection{Evaluation Setup }
\label{sec:setup}


\noindent\textbf{Evaluation Environments.}
Our evaluation environments include a ThinkPad P15 desktop and a Dell cloud server.
The ThinkPad desktop has an Intel i9-10885H vPro CPU (8 cores, 2.4 GHz), 128 GB RAM, 1TB storage, and an NVIDIA Quadro RTX 4000 Max-Q GPU, running Windows-11.
The Dell cloud server has an AMD EPYC-7643 CPU (88 usable cores, 2.3 GHz), 921 GB RAM, 12.8 TB storage, and 4 NVIDIA A100 GPUs with 80 GB VRAM each, supercharged by NVLink with the RHEL-8.6 operating system.

\paragraph{Model Execution}
We ran GPT-4 and ChatGPT on our binary code summarization datasets, completing the tasks in 20 and 6 days, respectively. 
GPT-4 tokenized all test samples into 309,401,126 tokens and generated responses totaling 19,678,448 tokens, summing up to 329,079,574 tokens, as reported in the first row of~\autoref{tab:model-stats}. 
For ChatGPT, the input samples were tokenized into 296,524,698 tokens, and it generated 16,338,012 tokens, resulting in a total of 312,862,710 tokens.
In terms of the cost, we spent \$10,462.74 and \$954.93 US dollars for querying GPT-4 and ChatGPT models, respectively.
For Llama 2 and Code Llama models, they tokenize our input samples into 736,413,994 tokens and generate 33,195,771, 29,699,869, 40,016,196, and 33,323,576 tokens as responses.
The execution of Llama models takes us 857 GPU hours to finish.
Finally, BinT5 produced 326,007,985 tokens, including 326,007,985 input and 8,456,320 output tokens within 16 GPU hours. 
To sum up, the model execution costs 11,418 US dollars and 873 NVIDIA A100 GPU hours.
\looseness=-1

\subsection{RQ1: Effectiveness}
\label{sec:eval-RQ1}
\input{sections/eval-RQ1}

\subsection{RQ2: LLM Comparison}
\label{sec:eval-RQ2}
\input{sections/eval-RQ2}

\subsection{RQ3: Various Binaries}
\label{sec:eval-RQ3}
\input{sections/eval-RQ3}

\subsection{RQ4: Other Factors}
\label{sec:eval-RQ4}
\input{sections/eval-RQ4}

%% file: sections/eval-RQ1.tex

\begin{figure}
  \centering
  \includegraphics[width=0.5\textwidth]{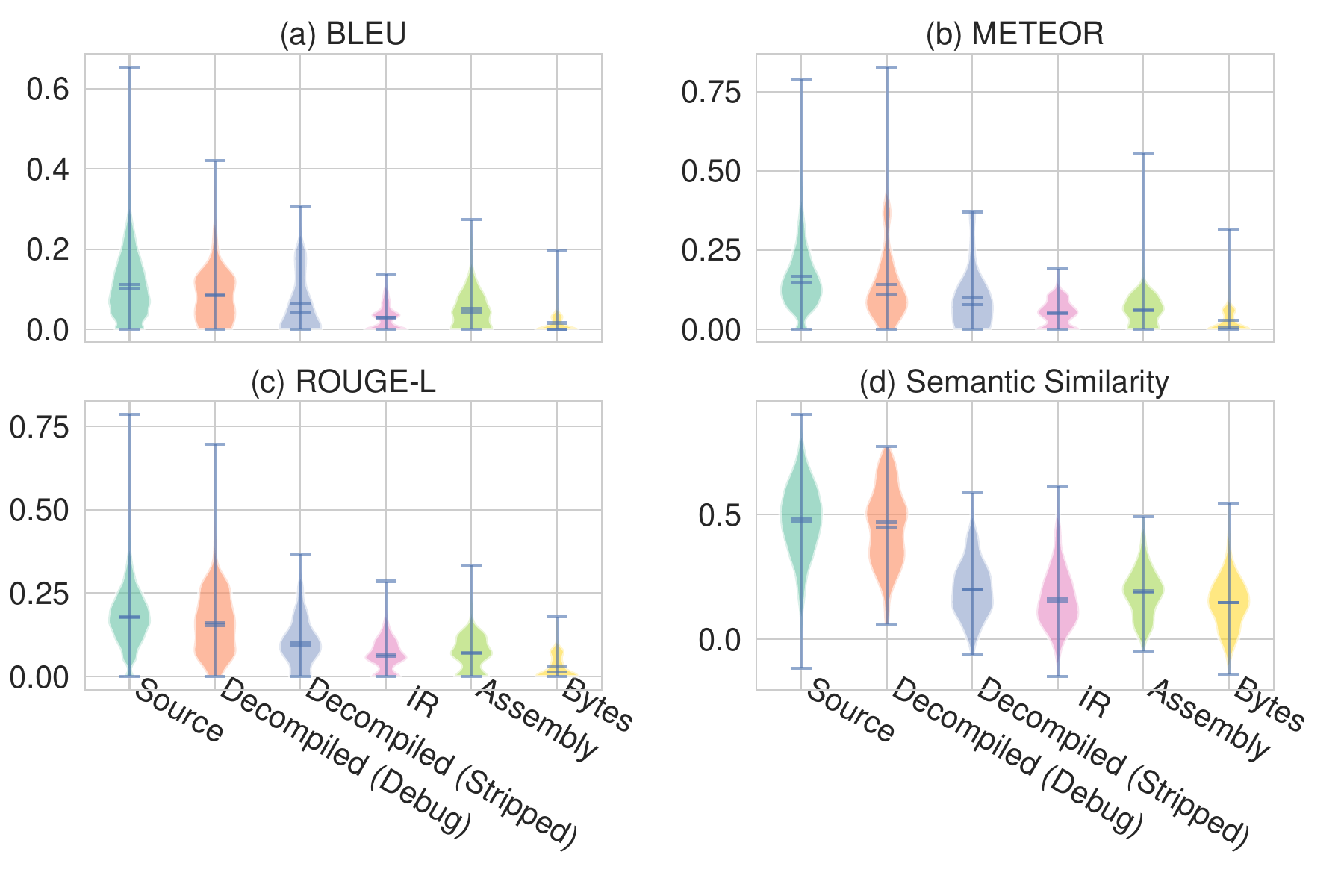}
  \caption{LLM Performance across Code Representations. For decompiled code, ``debug'' and ``stripped'' refer to these with and without debugging symbols. 
  }
  \label{fig:cross-representation}
\end{figure}

To answer \textbf{RQ1}, we first evaluate the performance of LLMs on different representations of binary code, including decompiled code, IR code, assembly code, and raw bytes along with source code, and present the results in \autoref{fig:cross-representation}, in which BinT5's performance is not included for a fair comparison, as it only accepts decompiled code.
Among the different code representations, we find that LLMs perform the best in source code, achieving an average semantic similarity score of 0.474.
For decompiled code, we observe a significant performance degradation when symbols are stripped from binaries, reducing the semantic similarity score from 0.449 to 0.202 (55.0\% decrease). \looseness=-1

\result{Stripping debugging symbols significantly loses decompiled code semantics by 55.0\%.}

Even with such a significant information loss, decompiled code without symbols still outperforms other binary code representations. 
Specifically, the average semantic similarity scores of assembly code, IR code, and raw bytes are 0.188, 0.185, and 0.118, respectively.
The same trend also appears in other evaluation metrics, \eg, the mean METEOR scores are 0.167, 0.141, 0.102, 0.052, 0.066, and 0.023 for source code, decompiled code (debug), decompiled code (stripped), IR code, assembly code, and raw bytes, respectively. \looseness=-1

\result{LLMs perform the best on decompiled code compared to IR code, assembly code, and raw bytes.}

\remove{After obtaining the statistical results, we also conduct an in-depth study to understand to which extent LLMs can comprehend binary code, as well as the nuanced interpretation of specific score outcomes. 
The result of this study is presented in Appendix \S\ref{case:study} for readers of interest. 
At a high level, our case studies of LLM-generated summaries provide additional confirmation of the deterioration of binary code semantics caused by symbol stripping (\fnumber{1}).
Furthermore, rather than elucidating high-level functionalities, LLMs tend to generate operational summaries for low-level binary representations, \ie, IR and assembly code (\fnumber{2}).
In the case of raw bytes, it is intriguing to note that the LLM-generated summaries closely resemble those of the assembly code, indicating the implicit code-lifting behavior (\fnumber{3}).
And, compared to exact matching-based metrics, our proposed semantic evaluation metric has been shown to be better suited for our task, as it effectively captures summary semantics (\fnumber{4}). 
}

\ignore{

Following the acquisition of statistical results, we perform a more in-depth study into the extent to which LLMs can comprehend binary code, as well as the nuanced interpretation of specific score outcomes.
For this, we perform a series of case studies on LLM-generated summaries. 
Our focus is on samples in which LLMs attain scores at the highest (100th percentile), median (50th percentile) and lowest (0th percentile) semantic similarity scores.
This investigative approach aims to provide us with distributional insights across extreme and median scenarios, thereby shedding light on the robustness of LLM in binary code comprehension.
For source code, our overarching observation is that LLM can capture most of its semantics (as presented in \autoref{tab:source-code-example} in Appendix).
For example, the summaries of source function 3 that obtain the median score show that the LLM has adeptly discerned the key semantics of ``skip characters'' and ``return the index''.
In addition, we observe a similarly good performance from summaries of decompiled code with debugging symbols (such as samples in \autoref{tab:decomp-debug-example}).
However, for decompiled code from stripped binaries, we find that LLMs cannot generate summaries explicitly matching the semantics of ground-truth summaries (as shown in \autoref{tab:decomp-stripped-example}).
For example, LLMs only manage to capture partial or elementary semantics of functions 2 and 3, despite their scores falling within the median range.
For function 2, the generated summary includes the semantics of ``file name canonicalization'' but it fails to include specific details of the ground truth, such as ``remov2 dots''.
For function 3, LLM only captures the ELF file format and loses all other semantic information.

\finding{Although LLMs excel at source code and decompiled code with symbols, the absence of debugging symbols leads to their ability to capture only partial or elementary semantics of decompiled code.}

For lower-level code, e.g., IR code and assembly code, we observe that LLMs predominantly focus on describing the low-level operations, without providing comprehensive high-level semantic summaries.
For instance, \autoref{tab:ir-code-example} presents the summaries and evaluation scores of the IR code.
Across samples that obtained the highest to lowest scores, LLM-generated summaries mostly describe the operations of data movement, arithmetic calculations (e.g., addition and subtraction), memory loading/writing, and data flow transformation (e.g., jump operation).
Similarly, we also observe such descriptions for assembly code as shown in \autoref{tab:assembly-example}.
Divergent slightly from the IR code summaries, we find that in the case of the highest score for assembly code, LLM can capture essential information related to the string data structure and the copy operation.
\finding{LLM-generated summaries fail to encapsulate the high-level semantics of IR and assembly code, instead, focusing on elucidating the low-level operations, e.g., data movement and arithmetic calculations.}

In the case of raw bytes, one would anticipate that LLMs generate summaries that describe a sequence of bytes. 
However, it is surprising to observe that LLM-generated summaries depict operations akin to those observed in IR code and assembly code (as shown in \autoref{tab:raw-bytes-example}), particularly for OpenAI models, i.e., ChatGPT and GPT-4.
For example, the summary of function 2 describes the stack frame operations, function calls, and arithmetic operations, which are not explicitly manifested in the raw byte input.
It appears that LLMs generate summaries by implicitly lifting the raw bytes into higher-level representations, such as assembly code that can represent the above-mentioned operations.

\finding{The LLM-generated summaries for raw bytes closely resemble those for assembly code, hinting at an implicit code lifting process performed by the LLMs.}




In addition to understanding the performance of LLMs in individual samples, we have also confirmed the benefits of using our proposed semantic evaluation metric.
Specificaly, as stated in \S\ref{sub:challenges}, metrics based solely on exact matching fail to provide a precise evaluation outcome, as they cannot encapsulate the semantic nuances of summaries.
For instance, for function 2 in \autoref{tab:source-code-example},  BLEU and METEOR scores are calculated as 0.
However, function 2's generated and ground truth summaries present close semantics, such as ``routine'' and ``function'', as well as ``a filename'' and ``the name of the file''.
These words/phrases are syntactically different, but semantically the same.
Moreover, its calculated scores of BLEU, METETOR, and ROUGE-L are lower than those of function 4, while function 2's summaries have shown closer semantics.
In contrast, our semantic evaluation metric computes semantic similarity at the semantic level, thus it generates fair scores that assign a higher score for closer semantics in function 2 compared to function 4.
Therefore, we argue that our semantic evaluation metric is more suitable for evaluating binary code summaries.

\finding{Our semantic evaluation metric can capture the essential semantics of binary code summaries, which is more suitable for our task than the exact matching-based metrics.}
}

%% file: sections/eval-RQ2.tex
\begin{figure}
  \centering
  \includegraphics[width=0.5\textwidth]{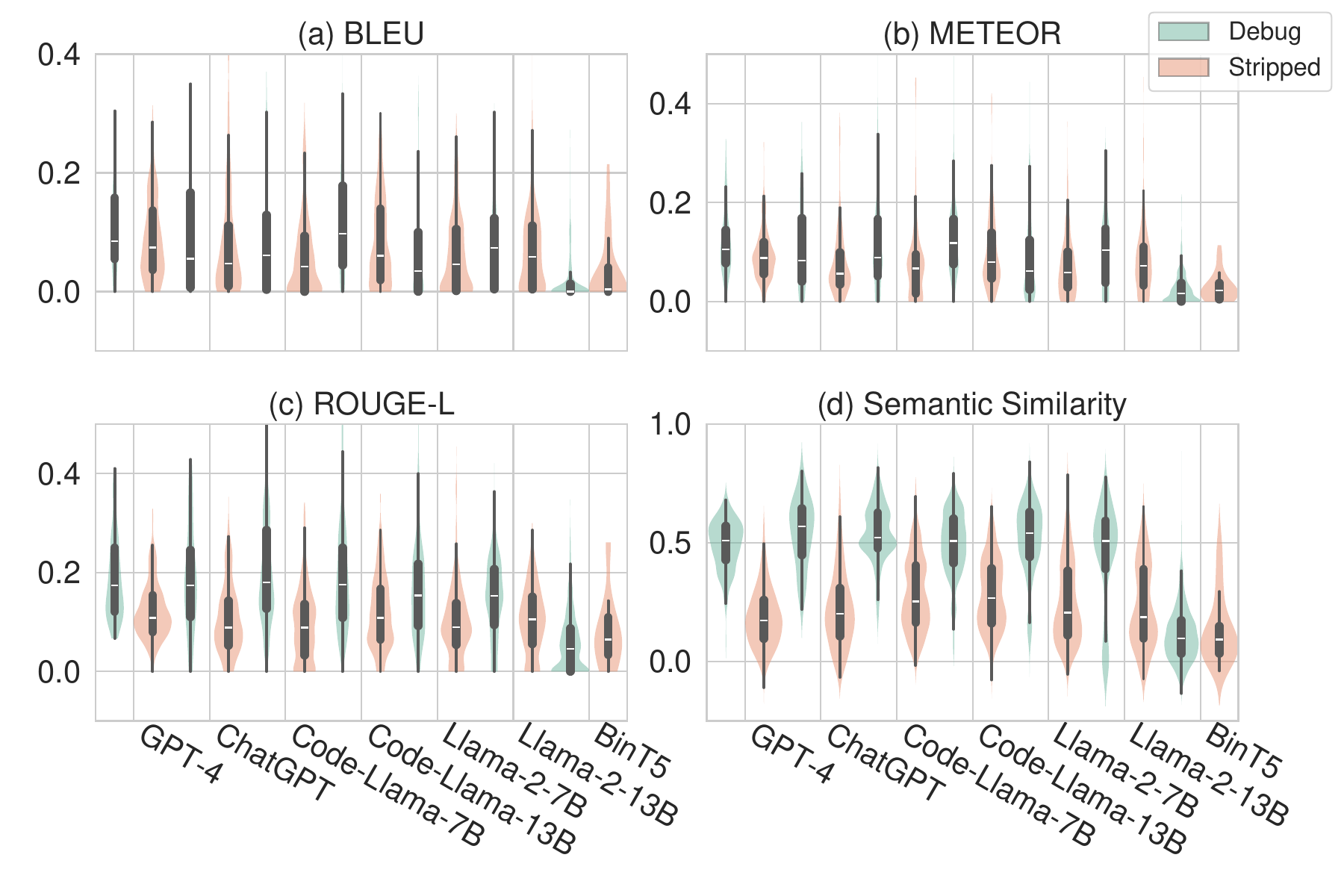}
  \caption{Performance across Different LLMs for Binaries with (Debug) and without (Stripped) Symbols
  }
  \label{fig:cross-model}
\end{figure}

To answer \textbf{RQ2}, we study individual performance for each of our target LLMs.
Having identified that decompiled code is the best for LLMs to understand binary code (\rnumber{2}), our report emphasizes the performance of LLMs on decompiled code, as shown in \autoref{fig:cross-model}.
For decompiled code with symbols, we find that ChatGPT performs the best, achieving an average semantic similarity score of 0.543. 
Meanwhile, for stripped binaries, Code Llama models outperform all other LLMs achieving 0.284 and 0.283 average semantic similarity scores for 7B and 13B models.
It is a surprise that GPT-4 is not the best-performing model. Therefore, we delve into a thorough examination of its results.
Overall, we observe that GPT-4 can understand binary code. 
However, it places a greater emphasis on the intricate and superfluous details, compared to ChatGPT, resulting in noisy summaries (\fnumber{5}).
\remove{
A more detailed analysis and specific examples are provided in Appendix \S\ref{sec:gpt-4-result}.
}

\result{ChatGPT and Code Llama perform the best for decompiled code with and without symbols, respectively, compared to GPT-4, Llama 2, and BinT5.}

Code Llama is fine-tuned on the code corpora from Llama 2~\cite{roziere2023code}. 
For decompiled code with symbols, Code Llama achieves 4.57\%, and 11.3\% better performance than Llama 2 on the 7B and 13B models.
For decompiled code from stripped binaries, Code Llama models further outclass Llama 2 model by 12.3\% and 22.0\% on the 7B and 13B models.
This outperformance of Code Llama demonstrates that Code Llama's fine-tuning can improve its capacity for binary code comprehension.
Meanwhile, BinT5 is the model fine-tuned from the CodeT5 model by retraining on decompiled code~\cite{al2023extending}.
Unfortunately, BinT5-generated summaries receive the lowest semantic similarity scores, \ie, 0.115 and 0.114, on binaries with and without symbols, compared to all other LLMs.
It is worth noting that for a fair and unbiased comparison among LLMs, we have excluded test samples that exist in BinT5's training dataset (See \S\ref{sec:bint5-outlier} for more discussions).
The marginal performance difference in binaries with and without symbols for BinT5 suggests that it may not fully capture the intricate semantics within binary code, even with symbols.
More importantly, we hypothesize that the different impact of fine-tuning in Code Llama and BinT5 may stem from disparities in their underlying base models, namely Llama 2 and CodeT5. \looseness=-1

\result{Code Llama, fine-tuned from  Llama 2, consistently outperforms Llama 2 (up to 22.0\% better scores). However, BinT5, fine-tuned from CodeT5, unfortunately, achieves the worst performance among all LLMs.}

\begin{table}[]
\caption{Efficiency of LLMs - Inference Time per Sample}
\resizebox{0.48\textwidth}{!}{
\begin{tabular}{cccccccc}
\toprule
\multirow{2}{*}{}                                                       & \multirow{2}{*}{\textbf{ChatGPT}} & \multirow{2}{*}{\textbf{GPT-4}} & \multicolumn{2}{c}{\textbf{Llama 2}} & \multicolumn{2}{c}{\textbf{Code Llama}} & \multirow{2}{*}{\textbf{BinT5}} \\ \cmidrule{4-5}\cmidrule{6-7}
&                                   &                                 & \textbf{7B}      & \textbf{13B}      & \textbf{7B}        & \textbf{13B}       &                                 \\\midrule
\textbf{\begin{tabular}[c]{@{}c@{}}Inference\\ Time (Sec)\end{tabular}} & 1.07                              & 3.1                             & 0.83            & 1.12             & 1.29              & 1.72              & 0.22    \\ \bottomrule                      

\end{tabular}
}
\label{tab:inference-time}
\end{table}

As listed in \autoref{tab:model-stats}, different LLMs have different model sizes and parameters, which can potentially lead to different efficiency.
Thus, we evaluate and compare the inference time cost per sample of our target models.
\autoref{tab:inference-time} presents the average inference time per sample for different LLMs.
Here, we have two different evaluation settings as stated in the model execution section of \S\ref{sec:setup}.
For ChatGPT and GPT-4, their execution relies on OpenAI API calls, making it challenging to determine the computational infrastructure that underpins their operation.
Their direct comparison shows that ChatGPT is 2.9$\times$ faster than GPT-4.
For other models, we have run them on our own cloud server which gives us a fair comparison.
Among these models, BinT5 is the fastest one among all models, \eg, 7.8$\times$ faster than Code Llama (13B).
We also find that Code Llama takes more time for inference compared to Llama 2, \eg, 1.5$\times$ more time on the 13B models.
In addition to the time cost, we have also observed the memory consumption difference among the open-source LLMs.
Specifically, the 7B and 13B Llama models occupy 42.6GB and 61.9GB of GPU VRAM, respectively, while the BinT5 model only takes 3.4GB of VRAM. \looseness=-1

\result{Compared to GPT-4, ChatGPT is 2.9$\times$ faster. For the other LLMs, BinT5 is the fastest model, while Llama 2 is (up to 1.5$\times$) faster than Code Llama.}

%% file: sections/eval-RQ3.tex
\begin{figure}
  \centering
  \includegraphics[width=0.5\textwidth]{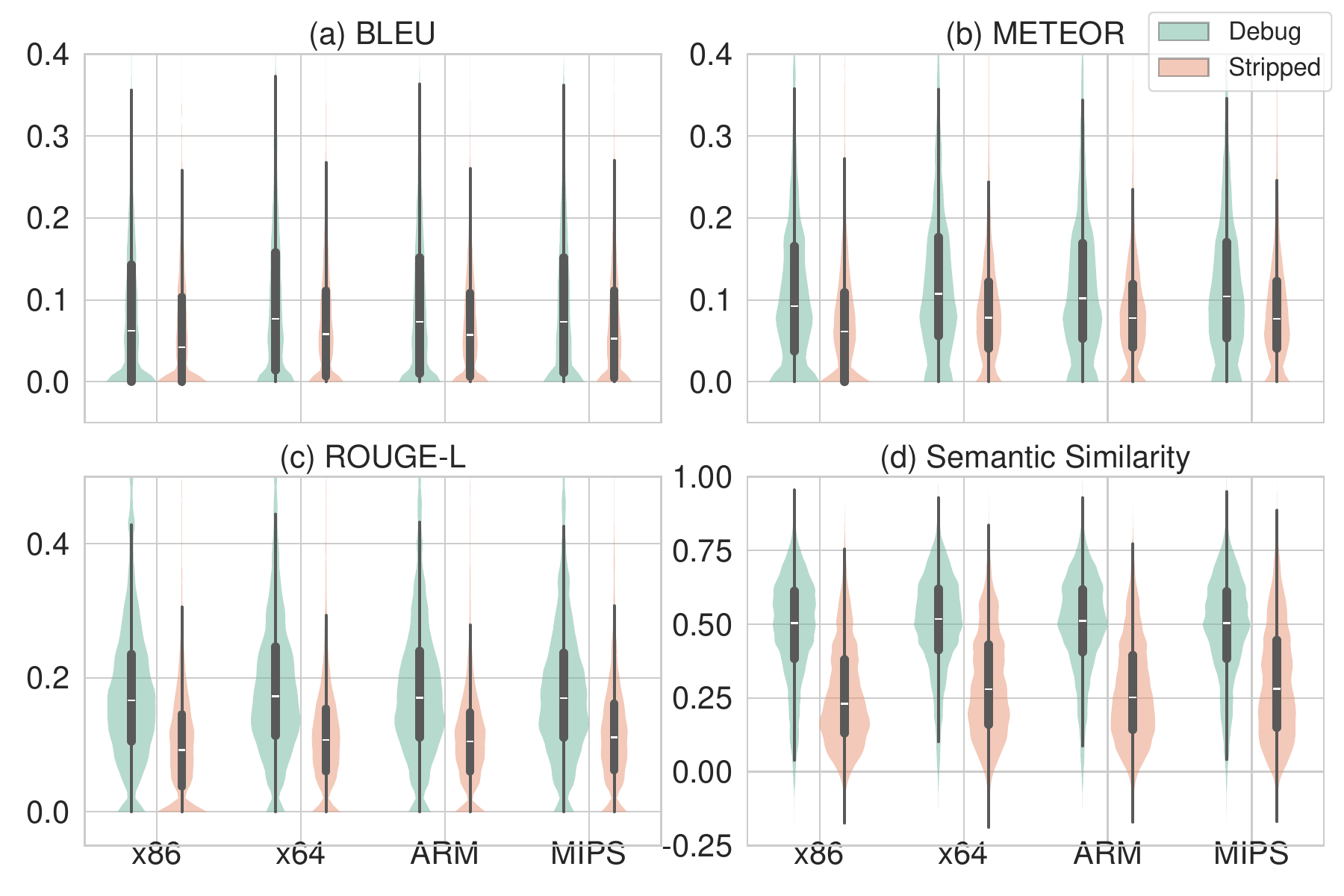}
  \caption{Performance across Computer Architectures for Binaries with (Debug) and without (Stripped) Symbols}
  \label{fig:cross-arch}
\end{figure}

To answer \textbf{RQ3}, we first study LLMs' performance on binaries across different computer architectures.
\autoref{fig:cross-arch} presents the performance on binaries across different architectures, \ie, x86, x64, ARM, and MIPS.
For binaries with symbols, the average semantic similarity scores for x86, x64, ARM, and MIPS binaries are 0.483, 0.503,  0.497, and 0.485, respectively.
The performance on x64 binaries outclasses these on x86 binaries by 4.14\%.
For stripped binaries, the similarity scores are 0.262, 0.299, 0.276, and 0.304 for the x86, x64, ARM, and MIPS binaries, respectively, in which LLMs' performance on MIPS binaries is 16.0\% better than these of x86 binaries. \looseness=-1

\result{
LLMs perform the best on x64 binaries with symbols and MIPS stripped binaries. The cross-architecture performance gap can be up to 16.0\%.
}

\begin{figure}
  \centering
  \includegraphics[width=0.5\textwidth]{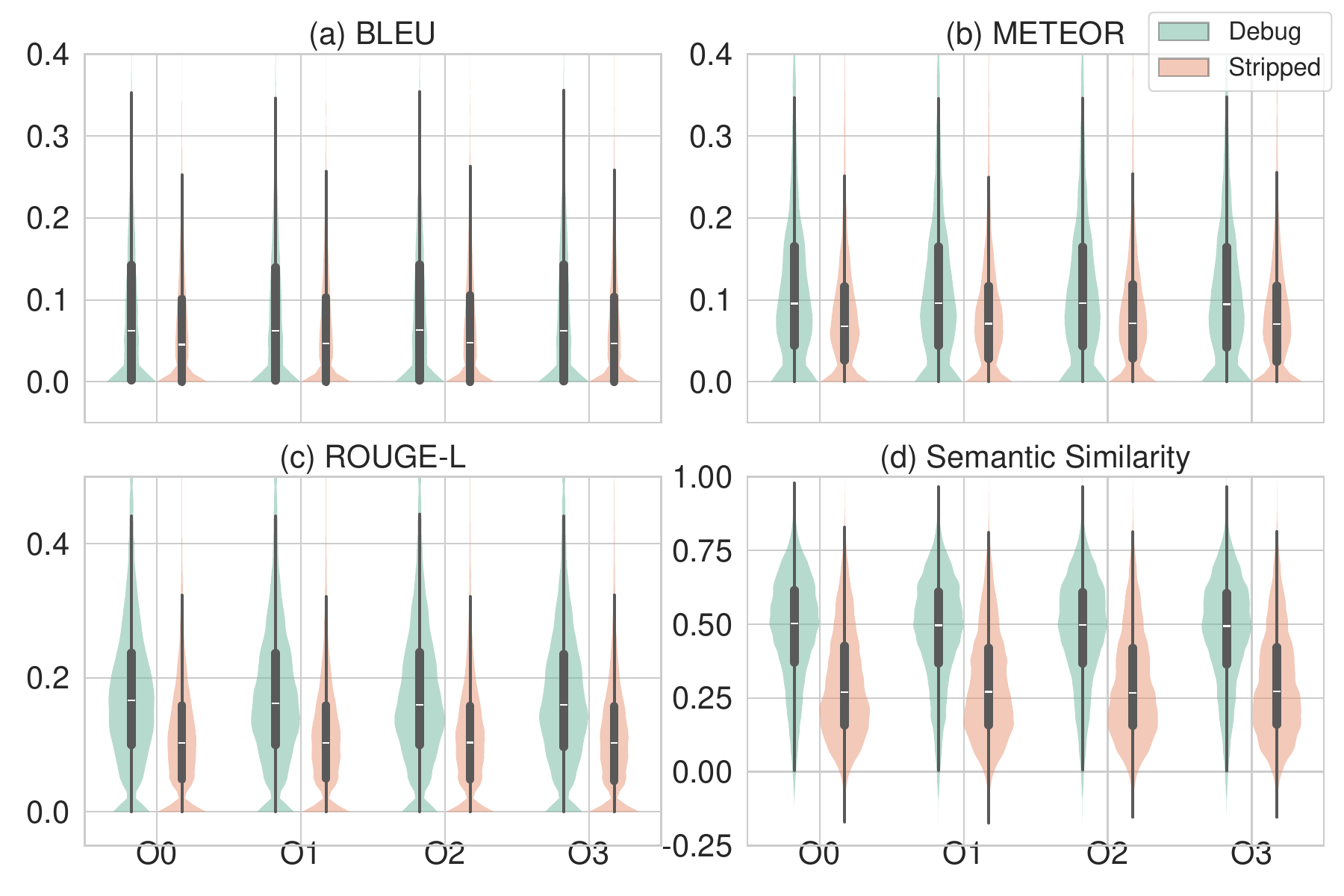}
  \caption{Performance across Optimization Levels for Binaries with (Debug) and without (Stripped) Symbols}
  \label{fig:cross-opt}
\end{figure}

In \autoref{fig:cross-opt}, we present the performance of LLMs when applied to binaries with O0, O1, O2, and O3 optimization levels. 
For binaries with symbols, LLMs attain average text similarity scores of 0.481, 0.477, 0.477, and 0.474 for O0, O1, O2, and O3 binaries, respectively.
Conversely, for stripped binaries, the average text similarity scores for these binary optimization levels are 0.269, 0.268, 0.266, and 0.270.
In both categories of binaries, we notice only a slight variance in performance across different optimization levels, with the performance gap between O0 and O3 binaries amounting to merely 1.47\%. \looseness=-1

\result{LLMs show marginal performance gap across different optimization levels, \eg, 1.47\% different text similarity scores between O0 and O3 binaries.}

%% file: sections/eval-RQ4.tex
\begin{figure}
  \centering
  \includegraphics[width=0.5\textwidth]{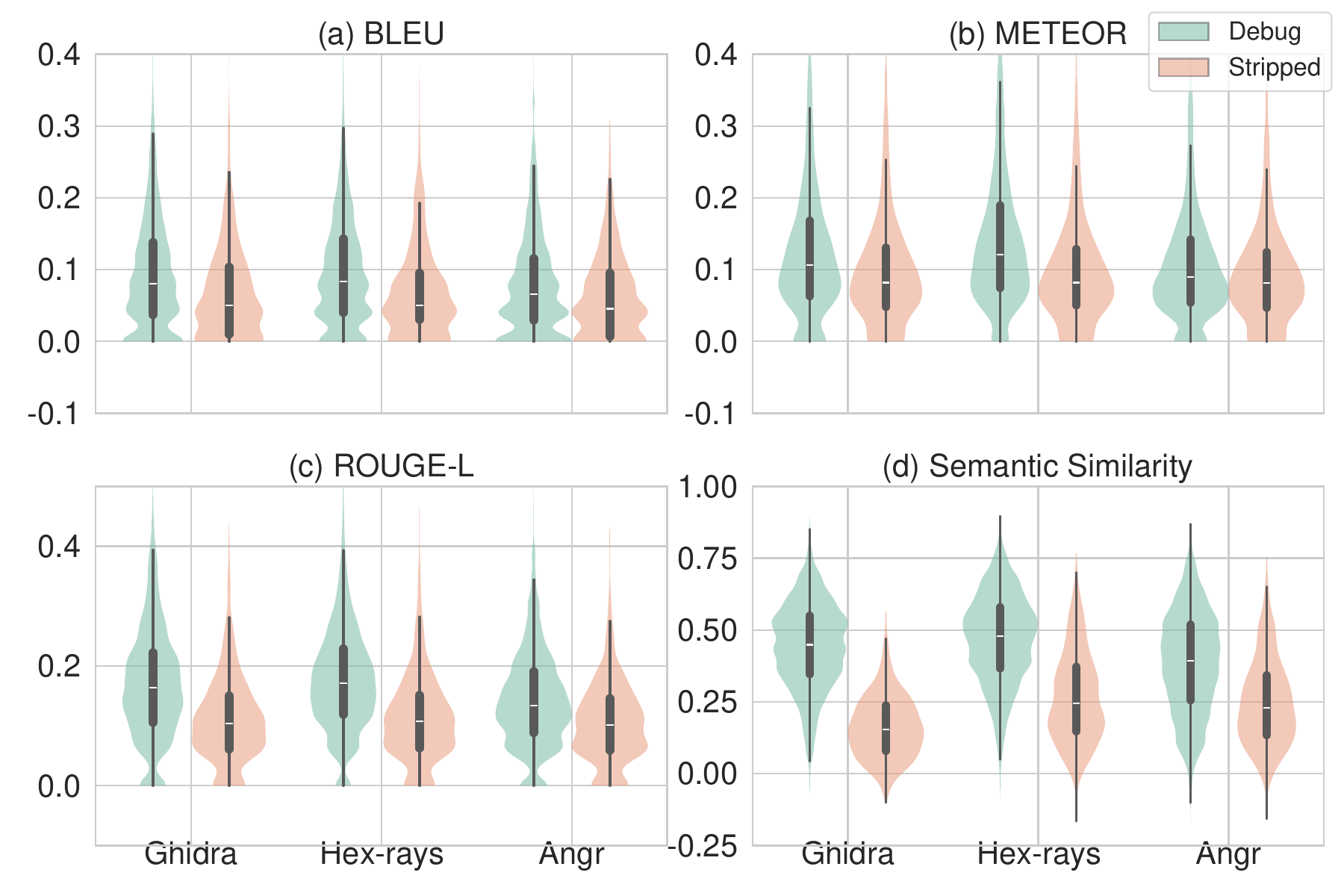}
  \caption{Performance across Different Decompilers for Binaries with (Debug) and without (Stripped) Symbols}
  \label{fig:cross-decompiler}
\end{figure}

Since we have identified that the decompiled code is the best for LLMs to understand (\rnumber{2}), next we undertake a further investigation of which decompiler generates the optimal output for code comprehension to answer \textbf{RQ4}.
\autoref{fig:cross-decompiler} presents the LLMs' performance on decompiled code generated from three different decompilers.
Overall, we find that Hex-rays consistently outperforms the other two decompilers for both binaries with and without symbols.
Specifically, for binaries with symbols, the output of Hex-rays achieves 5.41\% and 21.9\% higher semantic similarity scores than those of Ghidra and Angr.
For stripped binaries, it further outclasses Ghidra and Angr by 60.7\% and 19.8\%, respectively.
Among the decompilers, we observe that Angr is significantly slower than others.
For instance, Angr requires about 15$\times$ more decompilation time than Ghira with setting the \texttt{CFGFast} analysis for Angr, while Ghidra and Hex-rays consume the comparable time for the same task. \looseness=-1

\result{The decompiled code output of Hex-rays obtains the best scores compared to those of Ghidra and Angr, achieving up to  60.7\% outperformance. 
}

\begin{figure}
  \centering
  \includegraphics[width=0.5\textwidth]{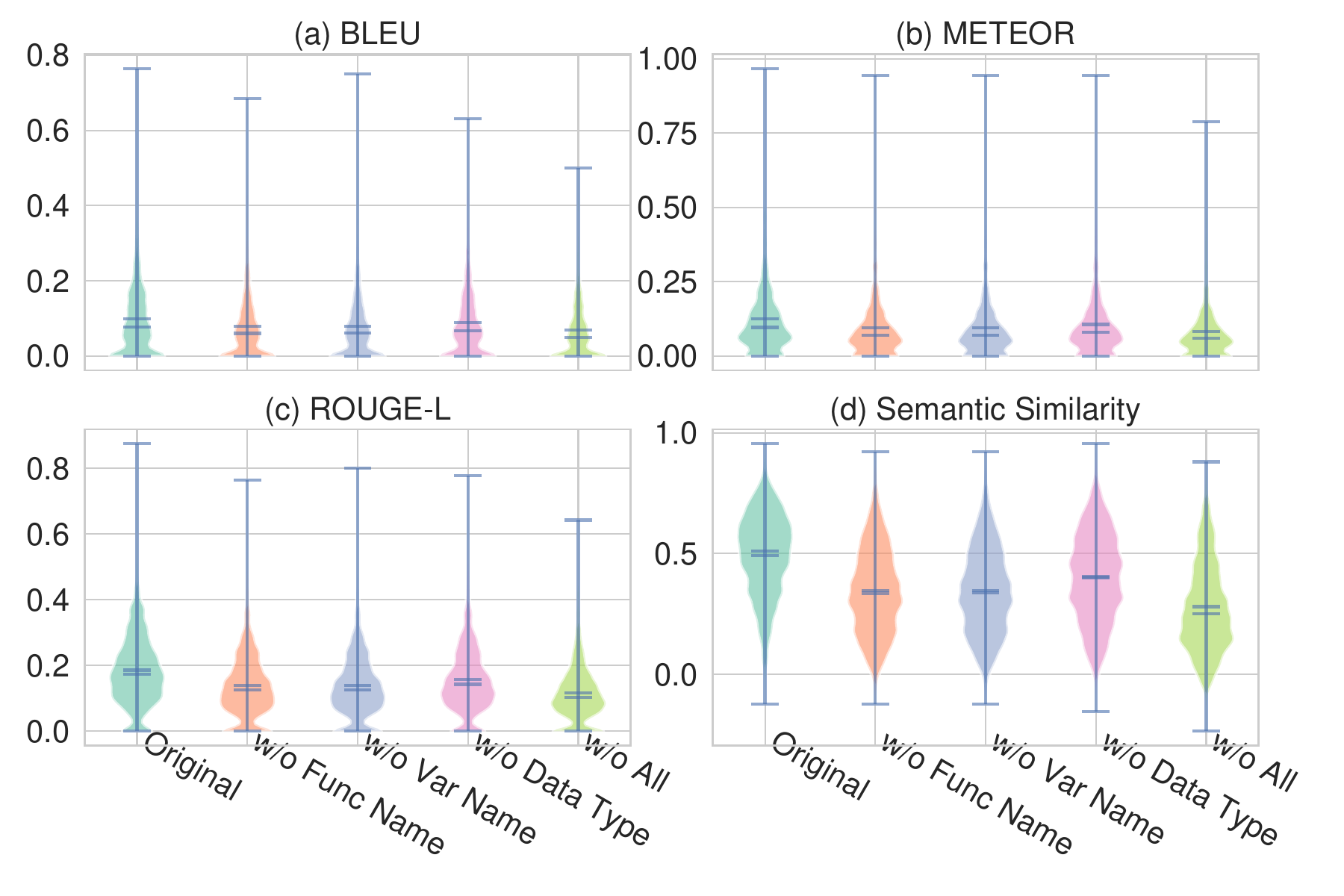}
  \caption{Semantic Impact of Stripping Function Names, Variable Names, and Data Types. We exclusively strip function names (w/o Func Name), variable names (w/o Var Name), and data types (w/o Type) from the decompiled code which originally has all symbols (Original). The baseline is decompiled code from stripped binaries (w/o All)}
  \label{fig:cross-stripping}
\end{figure}

Observing the performance decline from symbol stripping (\rnumber{1}), we delve deeper to discern which among the three symbol types—data types, variable names, and function names—most influence decompiled code semantics. 
\remove{Starting with the original decompiled code with all symbols as our baseline, we exclusively strip each symbol type by substituting them with non-informative placeholders (detailed in \S\ref{sec:symbol-stripping}).} 
\add{
Starting with the original decompiled code with all symbols as our baseline, we exclusively strip each symbol type by substituting them with non-informative placeholders. 
}
This produces three sets of modified decompiled code, with each having only one symbol type replaced while the rest remains intact. Additionally, we use the decompiled code from stripped binaries as an all-symbol-removed benchmark. 
\autoref{fig:cross-stripping} showcases the LLMs' performance across these variations. 
Stripping just the function names led to the most substantial drop in performance—a 30.3\% decrease in semantic similarity. 
Variable names and data types caused reductions of 30.0\% and 19.5\%, respectively. 
Moreover, we also find that the semantics of LLM-generated summaries can be manipulated by intentionally modified function names.
For this, we perform a case study and identify a vulnerability of LLM-generated summaries by function name manipulation (\fnumber{6}) in \S\ref{sec:summary-manipulation}.
\looseness=-1

\ignore{
Having noted the performance degradation caused by symbol stripping (\rnumber{1}), we further investigate which of the three types of symbols, i.e., data types, variable names, and function names, contribute the most to the decompiled code semantics.
For this, we start the experiment by considering the decompiled code with all symbols as the original code.
Subsequently, in order to assess the individual impact of each symbol type, we exclusively strip the corresponding symbols from the original code by substituting them with non-informative symbols (more details in \S\ref{sec:symbol-stripping}).
This process generates three sets of stripped decompiled code, in which we only replace symbols of each type in the decompiled code while keeping the rest of the decompiled code unaltered.
Finally, the decompiled code from stripped binaries serves as samples in which all symbols have been removed.
\autoref{fig:cross-stripping} presents LLMs' performance on the five sets of decompiled code.
Compared to the original binary code, only stripping the function names causes the most significant performance degradation, that is, 30. 3\% reduction in semantic similarity.
For variable names and data types, the decreases are 30.0\% and 19.5\%, respectively. 
}

\result{Function name contributes the most to decompiled code semantics. Merely stripping function names can reduce LLM's performance by 30.3\%.}

\remove{

}

\begin{figure}[]
  \centering
  \includegraphics[width=0.5\textwidth]{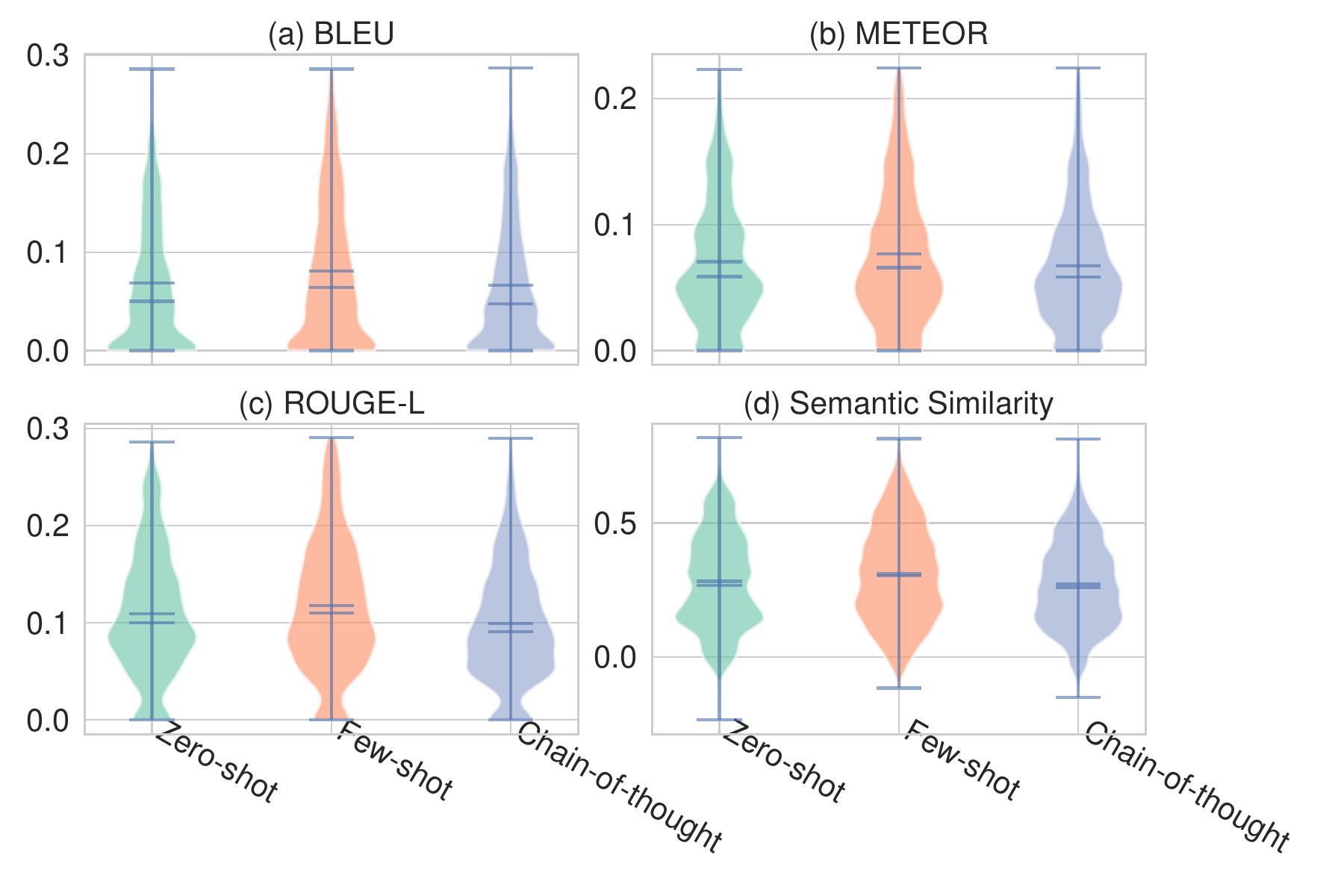}
  \caption{Performance on Different Prompts}
  \label{fig:cross-prompting}
\end{figure}    

\remove{
In our study, we predominantly employ zero-shot prompts.
However, we have also observed the use of advanced prompts, \eg, few-shot and chain-of-thought prompts, in NLP tasks~\cite{liu2023pre}; therefore, we investigate if these prompts can help LLMs in our task. Fundamentally, the few-shot prompts add demonstration examples (as shown in \autoref{fig:few-shot} in Appendix), \eg, pairs of binary code and corresponding summaries, to LLM input~\cite{brown2020language}.
For chain-of-thought prompting~\cite{wei2022chain}, we send two queries to LLMs (as exemplified in \autoref{fig:cot}).
In the first query, LLMs are first asked to explain the test code with the popular reasoning prompt ``Let’s think step by step''.
The resulting explanation is extracted and used to summarize the test code in the second query.
We provide more details of how we construct these prompts in \S\ref{sec:other-prompts}.
}
\add{
In our study, we predominantly employ zero-shot prompts.
However, we have also observed the use of advanced prompts, \eg, few-shot and chain-of-thought prompts, in NLP tasks~\cite{liu2023pre}; therefore, we investigate if these prompts can help LLMs in our task. Fundamentally, the few-shot prompts add demonstration examples, \eg, pairs of binary code and corresponding summaries, to LLM input~\cite{brown2020language}.
For chain-of-thought prompting~\cite{wei2022chain}, we send two queries to LLMs.
In the first query, LLMs are first asked to explain the test code with the popular reasoning prompt ``Let’s think step by step''.
The resulting explanation is extracted and used to summarize the test code in the second query.
}

\ignore{

Few-shot learning for LLMs was initially introduced to enhance the generalizability of GPT-3 for tasks beyond its primary domain~\cite{brown2020language}. 
This is achieved by imparting the model with contextual knowledge through in-context instructions, which is commonly referred to as zero-shot learning, or through demonstration samples, a technique known as few-shot learning.
To be more specific, few-shot learning entails the inclusion of a system instruction $\mathcal{I}$ along with either $n$ in-context demonstration pairs ($\mathcal{X}_{d}$, $\mathcal{Y}_{d}$), where $\mathcal{X}_{d}$ and $\mathcal{Y}_{d}$ represent sample inputs and outputs. This augmentation occurs prior to the introduction of the test input:

\begin{equation}
    x = \{\mathcal{I}; (x_d^{i}, y_d^{i})^{n}; x_{t}\}
\end{equation}
\noindent where the system instruction and demonstration samples are concatenated with the test input $x_{t}$ to form the query input $x$.
\autoref{fig:few-shot} presents the example prompt that we used to perform few-shot learning with the GPT-4 model.
Instead of using fixed demonstration examples, we randomly sample two pairs of example binary functions and summaries for each test binary code to avoid LLMs learning the unchanged pattern in the demonstration examples.
Chain-of-thought prompt is proposed to resolve complex reasoning problems by breaking them down into intermediate steps~\cite{wei2022chain}.
In the context of binary code understanding, we ask LLMs to perform an additional step of reasoning the input code semantics before generating the final code summaries.
For this, we first send one request asking LLMs to reason about the code semantics, in which we use the popular reasoning prompt ``Let’s think step by step''.
Upon receiving the response, we parse the generated code reasoning results and concatenate them into the second request inquiring about the code summaries.
\autoref{fig:cot} presents the sample of our chat-of-thought prompts for GPT-4 model.
To test these advanced prompt engineering techniques, we use the decompiled code samples from stripped binaries due to several reasons.
First, compared to decompiled code with symbols, our test decompiled code is more challenging to comprehend, demanding a deep understanding of the essential code semantics.
Second, the test-decompiled code contains less high-level semantic information, e.g., function and variable names, which can potentially distract LLMs in reasoning code semantics.
Finally, real-world binaries are commonly in stripped binary format, for which code summaries are more helpful.
For zero-shot learning, as exemplified in ~\autoref{fig:zero-shot}, we directly concatenate the prompt with the test code to generate the request.
}


\autoref{fig:cross-prompting} presents the performance of different prompts, where zero-shot, few-shot, and chain-of-thought prompts achieve average semantic similarity scores of 0.282, 0.312, and 0.271, respectively.
Compared to zero-shot prompts, few-shot prompts improve LLM performance by 10.6\%, while chain-of-thought reasoning does not help.
Beyond disparities in performance, we have also noted substantial variations in cost.
\add{Specifically, the median numbers of tokens for zero-shot, few-shot, and chain-of-thought prompts are 244, 1,763, and 944, respectively.}
\remove{
\autoref{fig:token-num-distribution} presents the length distribution of different prompts, where the median numbers of tokens for zero-shot, few-shot, and chain-of-thought prompts are 244, 1,763, and 944, respectively.
}
The 7.23$\times$ more tokens of few-shot prompts mean significantly higher computational cost.
For instance, since the use of ChatGPT and GPT-4 is changed based on token numbers, few-shot prompts would cost us 82,552 US dollars to finish the same amount of tokens listed in \autoref{tab:model-stats}.
Taking into account the scale of our evaluations, zero-shot learning can be a better solution based on the balance of performance and cost.

\result{Zero-shot prompts are better than few-shot and chain-of-thought prompts when jointly considering both performance and cost for our large-scale task.}


\remove{
\begin{figure}[t]
  \centering
  \includegraphics[width=0.4\textwidth]{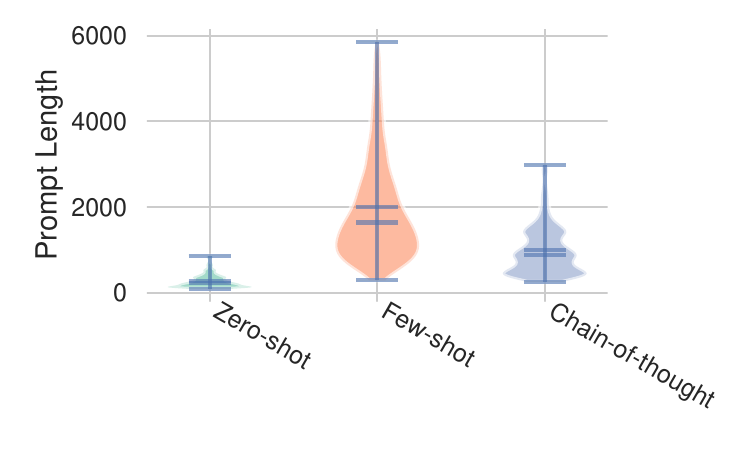}
  \caption{Prompt Length Distribution of Different Prompt Engineering Techniques}
  \label{fig:token-num-distribution}
\end{figure}
}

%% file: sections/lesson.tex

\noindent\textbf{Prioritizing Binary Code Comprehension is Paramount.}
The need for binary code summarization, albeit an area still in its infancy~\cite{al2023extending}, has rapidly grown, with rising demand for natural language summaries of binary code~\cite{virustotal-next-step-2023,google-cloud-blog-ai-insights,microsoft-security-copilot}. 
\remove{Notable entities like VirusTotal~\cite{virustotal-next-step-2023}, Google Cloud~\cite{google-cloud-blog-ai-insights}, and Microsoft Security Copilot~\cite{microsoft-security-copilot} are increasingly leveraging LLMs to understand malware/threats, aiming to produce ready-to-use summaries to security professionals.}  
Yet, binary code comprehension remains a formidable challenge, especially for stripped binaries. 
\remove{In our studies, LLMs either fail to grasp the complete semantics (\fnumber{1}) or generate summaries missing the nuances of high-level code semantics (\fnumber{2}).} 
The multifaceted nature of binaries, along with their diverse representations, compounds this challenge. 
For instance, LLM performance varies across binary representations (\rnumber{2}) and is sensitive to decompiler outputs (\rnumber{8}). Additionally, differing compilation settings, such as computer architectures, can influence LLM outcomes (\rnumber{6}). In navigating this intricate landscape, our findings offer pivotal insights that can guide future endeavours in comprehensively understanding binary code.

\ignore{
\smallskip
\noindent\textbf{Lesson 1: We need to focus efforts on binary code comprehension}: 
While binary code summarization is understudied~\cite{al2023extending}, there is a growing demand to produce natural language summaries for binary code~\cite{virustotal-next-step-2023,google-cloud-blog-ai-insights,microsoft-security-copilot}.
VirusTotal~\cite{virustotal-next-step-2023}, Google Cloud~\cite{google-cloud-blog-ai-insights}, and Microsoft Security Copilot~\cite{microsoft-security-copilot} have been seeking to leverage LLMs to identify and summarize malware/threats for the benefit of security analysts. 
However, the challenge of this task is far from being resolved, particularly for stripped binaries.
We have found that LLMs either fall short in capturing the complete semantics (\fnumber{1}) or produce summaries that lack high-level code semantics (\fnumber{2}).
Meanwhile, the diverse binaries and binary representations further make this problem even more challenging.
For example, given different binary representations, LLMs have shown various performance (\rnumber{2}).
Even for the same decompiled code representation, the output of different decompilers leads to different LLM performance (\rnumber{8}).
The different compilation settings of binaries, e.g., computer architectures, can also affect LLMs (\rnumber{6}).
While facing these challenges, we believe that our measurement results and findings contribute unique and valuable insights toward addressing these issues.
}




\smallskip
\noindent \textbf{Harnessing LLMs Requires Further Advancements and Explorations.}
\remove{Machine learning models are extensively utilized in binary analysis, yet many face challenges in generalizing to unseen samples~\cite{jin2022symlm,chen2022augmenting}. 
In contrast, LLMs like ChatGPT and GPT-4 boast superior generalizability—crucial for the vast array of real-world binaries. 
However, these promising models come with limitations. 
They are resource-intensive, with GPT and Llama models demanding expensive computational resources and specialized hardware accelerators (\S\ref{sec:eval-RQ2}). Potential solutions may lie in model distillation and hardware acceleration techniques~\cite{deng2020model}, especially as smaller models have proven to be more resource-friendly (\rnumber{5}). The black-box nature of LLMs presents another challenge, making it tricky to decipher their exact behaviors. Yet, our work suggests that refining and optimizing prompts might be a pathway forward, as highlighted by our prompt evaluations (\autoref{tab:top-prompts} and \rnumber{10}). Additionally, while fine-tuning LLMs can enhance binary code comprehension, selecting the right base models is essential, as underscored by our findings in \rnumber{4}. In essence, while LLMs hold immense promise in binary code analysis, their effective deployment hinges on computational challenges, requiring selecting the correct models and inputs. }
\add{Machine learning models are extensively utilized in binary analysis, yet fail to unseen samples~\cite{jin2022symlm,chen2022augmenting}. 
In contrast, LLMs boast superior generalizability for real-world binaries. 
However, LLMs are resource-intensive, demanding expensive computational resources and specialized hardware accelerators (\S\ref{sec:eval-RQ2}). Potential solutions may lie in model distillation and hardware acceleration techniques~\cite{deng2020model}, especially as smaller models have proven to be more resource-friendly (\rnumber{5}). 
Meanwhile, the black-box nature of LLMs presents another challenge, making it tricky to decipher their exact behaviors. Yet, our work suggests that refining and optimizing prompts might be a pathway forward (\rnumber{10}). Additionally, while fine-tuning LLMs can enhance binary code comprehension, selecting the right base models is essential (\rnumber{4}). In essence, while LLMs hold immense promise in binary code analysis, their effective deployment hinges on computational challenges, requiring selecting the correct models and inputs. }
\looseness=-1

\ignore{
\noindent \textbf{Lesson 2: We need to further improve and explore LLMs}:
Machine learning models are extensively employed in binary analysis; however, they often exhibit limited generalizability~\cite{jin2022symlm,chen2022augmenting}.
In contrast, LLMs like ChatGPT and GPT-4 have demonstrated remarkable generalizability, a valuable attribute for our task, given the diverse real-world binaries.
However, it is imperative to acknowledge that these LLMs come with their own set of limitations, including the high computational cost and their black-box nature. 
First, LLMs such as GPT and Llama models can be extremely expensive, often necessitating decent hardware accelerators (\ref{sec:eval-RQ2}).
For this, model distillation and hardware acceleration~\cite{deng2020model} might be useful, considering that smaller models are computationally lightweight and efficient (\rnumber{5}).
Furthermore, the black-box nature of LLMs makes it challenging to determine the optimal input of binary code.
Exploring different prompts by prompt synthesis and optimization can be useful based on our prompt evaluation results (\autoref{tab:top-prompts} and \rnumber{10}).
Finally, fine-tuning LLMs (even on source code) can improve the understanding of binary code, but the base models should be carefully selected (as found in \rnumber{4}).

}

\smallskip \noindent \textbf{Augmenting Decompilers is Crucial for Improved Binary Comprehension.}
Our studies indicate that decompiled code proves most intuitive for LLMs (\rnumber{2}).
Enhancing decompilers, particularly for stripped binaries, cannot be overstated. 
Our findings suggest that binary stripping can greatly diminish the semantics of decompiled code (\rnumber{1}). 
Furthermore, aside from BinT5~\cite{al2023extending}, all other LLMs exhibit a significant performance gap for binaries with and without symbols (\autoref{fig:cross-model}). 
This leads to LLMs' inability to fully capture the semantics of stripped binary functions (\fnumber{1}). 
A promising approach to strengthen decompilers can be symbol recovery~\cite{chen2022augmenting,jin2022symlm,banerjee2021variable}, as depicted in \autoref{fig:cross-stripping}. 
Our research pinpoints function name prediction~\cite{jin2022symlm,david2020neural,patrick2023xfl} as particularly influential, given its paramount contribution to the semantics of decompiled code (\rnumber{9}). 
The consistency check of binary function names is also important, particularly because LLM-generated summaries are susceptible to malicious manipulation through intentional alterations of function names (\fnumber{6}).
Currently, Hex-Rays stands out as a leading decompiler, consistently producing output that LLMs perform the best on (\rnumber{8}). Moving forward, we believe that the insights from our study can serve as invaluable guidance for improving decompilers. \looseness=-1

\ignore{
\noindent \textbf{Lession 3: We need to enhance decompilers}:
While we have identified that decompiled code is the most effective for LLMs to comprehend (\rnumber{2}), it is imperative to underscore the significance of improving decompilation output for stripped binaries based on the following results.
First, binary stripping can significantly reduce the semantics of decompiled code (\rnumber{1}). 
Furthermore, there is a notable performance discrepancy between binaries with symbols and stripped binaries among all LLMs, except for BinT5 (\autoref{fig:cross-model}). 
Consequently, LLMs fall short in capturing the comprehensive semantics preserved within stripped binary functions (\fnumber{1}).
A potential direction of enhancing decompilers is symbol recovery~\cite{chen2022augmenting,jin2022symlm,banerjee2021variable}, which is critical to helping LLMs understand stripped binaries as shown in \autoref{fig:cross-stripping}.
Function name prediction~\cite{jin2022symlm,david2020neural,patrick2023xfl}, in particular, stands out as a valuable technique, as it contributes the most to decompiled code semantics (\rnumber{9}).
At present, some decompilers, such as Hex-Rays, have emerged as frontrunners in producing superior output for LLMs (\rnumber{8}). 
In the future, we believe that the insights from our measurement study can serve as invaluable guidance for improving decompilers.
}



%% file: sections/discussion.tex

\noindent \textbf{Scope of Study and Binary Diversity.}
In this study, we have performed an extensive study of LLM's capabilities for binary code comprehension, assessing over 557K binary functions across different architectures and optimization levels. However, real-world binaries extend beyond the confines of our dataset, encompassing domain-centric binaries like IoT firmware and malware. Such binaries often possess unique semantics which might be further complicated by code obfuscations. Moreover, while our study has assessed specific binary code representations, it does not encompass all possible representations, such as those generated by other decompilers or IR code generators. Delving into these areas presents a promising avenue for future research. \looseness=-1

\paragraph{Emergence of New LLMs and Prompt Techniques}
Our study evaluated four leading LLMs recognized for their exceptional performance across numerous tasks~\cite{roziere2023code,openai2023gpt}, as well as the binary code summarization model, BinT5~\cite{al2023extending}. Still, with academia and industry ceaselessly advancing, newer LLMs are continually being introduced, and these might be applicable for our task. Additionally, while our research employed a prompt synthesis and optimization method yielding 320 prompts (\S\ref{sec:setup}), there remains potential for discovering even more effective prompts and refining prompt engineering methodologies, like gradient-based prompt tuning~\cite{liu2023pre}.

\paragraph{Code Summary and Semantic Representations}
In our research, developer-written code comments served as our gold standard for code summaries. 
Alternative representations for these summaries exist as well, such as analysis logs or system calls~\cite{hao2023syzdescribe,pan2023automated}. 
However, these representations are artificial, generated by third parties instead of the code creators, and might not always align semantically with the original code. 
We observe that developers' comments present a more intuitive option. 
This aligns with conventional methodologies utilized in prior studies~\cite{feng2020codebert,husain2019codesearchnet,al2023extending}. 
Similar to well-known code summary datasets, \eg, CodeSearchNet~\cite{husain2019codesearchnet}, our dataset may also include semantic inconsistent noise~\cite{chen2021my} introduced by developers.
While mitigating this noise is still an open question, this would be a worthwhile focus for subsequent research.
\looseness=-1

\ignore{
\paragraph{A best-effort and large-scale study}
We have performed a large-scale and comprehensive study on this task.
Our evaluations cover over 557K binary functions with different binary code representations from binaries across four architectures and four optimization levels.
However, the real-world binaries are not limited to those in our data set.
These binaries, e.g., IoT firmware and malware, often feature domain-specific semantics, which may be intertwined with code obfuscations.
Furthermore, it's worth noting that binary code representations extend beyond the scope of those examined in our paper, including those from other decompilers and IR code generators.
We believe that evaluating the other binaries and representations can be an interesting future work.

\paragraph{More LLMs and Prompts}
We have tested four popular LLMs that exhibit state-of-the-art performance in many real-world tasks~\cite{roziere2023code,openai2023gpt}, along with the existing binary code summarization model BinT5~\cite{al2023extending}.
However, it's worth noting that an increasing number of LLMs are continually being developed and released by both academic and industry sources. 
These newly emerging LLMs could potentially find relevance and application in our task.
In addition, we have proposed the prompt synthesis and optimization approach to generate 320 prompts (\S\ref{sec:setup}) and selected the best in our evaluations.
However, there could be other effective prompts and advanced prompt engineering techniques, such as gradient-based prompt tuning~\cite{liu2023pre}, for our task.



\paragraph{Alternative Code Summaries}
In this paper, we have adopted developer-written code comments as the gold standard for code summaries. 
It's worth noting that the code summaries be represented in alternative formats, such as analysis logs and system calls~\cite{hao2023syzdescribe,pan2023automated}. 
However, in contrast to code comments that effectively convey the core code semantics, these alternative code summaries are artificial and generated by third parties. 
Consequently, they may exhibit a semantic shift in comparison to the original code semantics.
Human language-based code comments can be more accessible and easier for security analysts to understand, compared to alternative forms.
Therefore, we opt to use code comments as ground truth, following the common practice of existing research~\cite{feng2020codebert,husain2019codesearchnet,al2023extending}.
}

%% file: sections/related.tex

\noindent \textbf{LLM for Binary Reverse Engineering.}
Binary reverse engineering is labor-intensive and suffers from the absence of high-level semantics, especially for stripped binaries.
Existing LLM-based solutions follow the pretraining-finetuning paradigm to learn binary semantics from large binary corpora and transfer the learned knowledge to downstream tasks, \eg, function similarity detection~\cite{li2021palmtree,pei2020trex,wang2022jtrans}, variable name and type inference~\cite{pei2021stateformer,chen2022augmenting,banerjee2021variable}, and function name prediction~\cite{jin2022symlm,david2020neural,patrick2023xfl}.
For example, Trex~\cite{pei2020trex} pretrains roBERTa on the assembly code and execution trace to detect function similarity.
SymLM~\cite{jin2022symlm} predicts function names by learning calling context and execution behavior.
Unlike existing works that focus on BERT-level LLMs, we focus on generative LLMs that exhibit significantly better generalizability and emergent abilities~\cite{wei2022emergent}. \looseness=-1

\paragraph{Automated Code Summarization}
Automated code summarization aims at articulating code fragments, typically methods or functions,  using natural language summaries~\cite{shi2022evaluation}.
Recently, the predominant focus of this research has been on source code~\cite{guo2020graphcodebert,wang2021codet5,feng2020codebert}
As an example, CodeT5~\cite{wang2021codet5} pretrains the T5 model with semantics-enriched code features, \eg, identifiers, to summarize code.
GraphCodeBERT summarizes binary code semantics by modeling the data flow with the BERT model~\cite{guo2020graphcodebert}.
In addition to source code, there are also research efforts on binary code.
For example, BinT5 finetunes CodeT5 on decompiled code to generate natural language descriptions~\cite{al2023extending}.
Different from existing research, we are the first to perform a large-scale and comprehensive study of different LLMs on the binary code summarization task.

\ignore{
\paragraph{Automated Code Summarization}
Automated code summarization has attracted long-term attention in the software engineering community.
For instance, CodeBERT is pretrained on the BERT model for source code summarization and generation~\cite{feng2020codebert}.
Different from the encoder-only model of CodeBERT, CodeT5 is pretrained with the encoder-decoder model, i.e., T5, for Code Understanding and Generation~\cite{wang2021codet5}.
Codex, a GPT--3 based LLM finetuned on public source code repositories, have shown promising results in source code understanding tasks~\cite{chen2021evaluating}, which has recently been deprecated and replaced by GPT-4~\cite{openai-deprecations}.
Recently, MetaAI released the Code Llama model which is finetuned on the Llama 2 model, which also exhibit strong code understanding capacities~\cite{roziere2023code}.
In addition to source code summarization, BinT5, the only binary code summarization model to the best of our knowledge, is finetuned from CodeT5 on the Ghidra decompilation output~\cite{al2023extending}.
}

\ignore{
\paragraph{Prompt Engineering}
It's worth noting that, similar to existing applications of ChatGPT in software engineering tasks \cite{sobania2023analysis, tian2023chatgpt, sobania2023analysis}, our primary emphasis lies in the design of optimal zero-shot prompts for LLMs, rather than delving into advanced prompt engineering techniques like few-shot instruction tuning \cite{honovich2022instruction} and chain-of-thought. 
The advanced prompt engineering techniques has several limitations.
First, ChatGPT and GPT-4 are charged per token, thus the advanced prompt engineering techniques will dramatically increase our experiment cost consider the scale of our experiments.
For example, for the few-shot prompting with two demonstration samples, the total cost will increase by at least three times.
Second, the advanced prompt engineering techniques can bring the biases, resulting in unfair comparison, into our evaluations as they have exhibited different performance on different tasks~\cite{suzgun2022challenging}, such as chain-of-thought may benefit decompiled code summarization more than assembly code summarization.
Although we do explore the effectiveness of some advanced prompting approaches in our ablation study in \S\ref{}, we leave the exploration of these techniques for future work.
}


%% file: sections/conclusion.tex
We have presented a large-scale and comprehensive study of how LLMs can understand binary code semantics. 
We built \sysname, a comprehensive benchmark with an expansive dataset with over 557K binary functions, spanning various code representations, computer architectures, and optimization levels.
We construct an extensive binary code summarization dataset, with over 557K binary functions in different binary code representations across different computer architectures and optimization levels.
We designed the novel in-context prompt synthesis and optimization techniques for optimal prompt generation.
We also devised a new semantic evaluation metric to measure code summaries.
Our rigorous evaluations have resulted in 10 empirical results and 6 findings, which provide nuanced insights into both LLMs and binary code, serving as a reference for future research in binary code comprehension. \looseness=-1


%% file: sections/appendix.tex
\subsection{Source Projects}

\input{tables/src-project-lines}

\autoref{tab:source-projects} presents the 44 open-source projects that we used to compile binaries.

\subsection{Cross-compilation and Stripping}
\label{sec:cross-compiler}

Our compilation and stripping processes are performed on a 64-bit Ubuntu machine. 
Compiling and stripping binaries with computer architectures different from the host machine, \ie, ARM and MIPS, requires cross-compilers and cross-stripping tools.
For this, we have used cross-compilers, \texttt{arm-linux-gnueabihf-gcc} and \texttt{mipsel-linux-gnu-gcc} for ARM and MIPS binaries.
For binary stripping, we use the \texttt{strip} command for x86 and x64 binaries. For ARM and MIPS binaries, we leverage \texttt{arm-linux-gnueabihf-strip} and \texttt{mipsel-linux-gnu-strip}.

\subsection{Binary Functions with Ground-truth Summaries}

\input{tables/binary-function-distribution}

\autoref{tab:labeled-binary-functions} presents the distribution of binary functions across computer architectures and optimization levels with ground-truth summaries, extracted by methods defined in \S\ref{sec:dataset}.

\subsection{Evaluation Metrics for N-gram Matching}
\label{sec:exact-matching-metrics}
In this paper, we report three widely used n-gram matching-based evaluation metrics, \ie, BLEU, METEOR, and ROUGE-L. \looseness=-1

\paragraph{BLEU}
Bilingual Evaluation Understudy (BLEU) \cite{papineni2002bleu} is a commonly employed metric for assessing the quality of generated texts versus reference texts. 
It is a variant of precision metrics and quantifies similarity by computing the n-gram precision between a generated summary and a reference summary, with a penalty for excessively short lengths. 
The BLEU score is computed as follows:

\begin{equation}
\text{BLEU} = \text{BP} \times \exp\left(\sum_{n=1}^{N} \frac{1}{N} \log \text{p}_n\right)
\end{equation}

\noindent where $\text{BP}$ represents the brevity penalty, $N$ signifies the maximum order of n-grams taken into account, and $\text{p}_n$ denotes the precision associated with n-grams. 
In this paper, we report the BLEU-1 score which is calculated on the unigram matching results. We have also calculated BLEU-2, BLEU-3, and BLEU-4, but their scores are 0 for most samples.

\paragraph{METEOR}
Metric for Evaluation of Translation with Explicit ORdering (METEOR) \cite{banerjee2005meteor} is proposed to improve the measurement of text ordering.  
In the context of comparing a pair of summaries, METEOR establishes a word alignment between them and subsequently computes similarity scores.
Formally, METEOR is calculated by:
\begin{equation}
    \text{METEOR}= (1-\gamma\times f^{\beta})\times\frac{P\times R}{\alpha\times P + (1-\alpha)\times R}
\end{equation}

\noindent where $P$ and $R$ are the precision and recall of the mapped unigrams in generated summaries and summary references. $f$ is the fragmentation coefficient. 
The penalty parameters $\alpha$, $\beta$, and $\gamma$ come with default values of 0.9, 3.0, and 0.5, respectively.

\paragraph{ROUGE-L}
ROUGE-L is a variant of Recall-oriented Understudy for Gisting Evaluation (ROUGE), calculated based on the longest common subsequence (LCS) between the pair of texts.
Specifically, ROGUE-L is calculated by:
\begin{equation}
F_{lcs} = \frac{(1 + \beta^2)\times R_{lcs}\times P_{lcs}}{R_{lcs} + \beta^2\times P_{lcs}}
\end{equation}

\noindent where $P_{lcs}$ and $R_{lcs}$ are the precision and recall of the LCS between generated summaries and summary references. $\beta$ governs the relative significance of precision and recall, which is commonly set to 1.2.

\subsection{Prompt Evaluation Results}
\label{sec:prompt-evaluation}
\input{tables/selected-prompt}
\autoref{tab:top-prompts} presents the top 40 prompts and their evaluation results, which are generated by our in-context prompt synthesis and optimization approach (proposed in \S\ref{sec:prompt-engineering}).
As mentioned in \S\ref{sec:implementation}, we first generate 320 prompts by GPT-4, and then evaluate each of them on 1000 binary function samples that are randomly selected from our binary summarization dataset.
To generate the 320 prompts, we chose to use GPT-4 because these tasks essentially involve test generation, and GPT-4 has shown advanced performance in this area according to OpenAI's documentation \cite{openai2023gpt}.
Based on the semantic similarity score, we select the best prompt in our subsequent evaluation.
In addition to the prompt text, we also added instructions to limit the summary length as discussed in \S\ref{sec:testing-eval}.
To ease the process of retrieving summaries, we get structured summary output by adding the following text to LLM inputs, such as 
``Input decompiled code:\textbackslash n<CODE>\textbackslash n Function Summary:'' for decompiled code summarization, in which <CODE> is filled with test decompiled code.
For the GPT-4, ChatGPT, Llama 2, and Code Llama models, we use the same prompt for a fair comparison.
For BinT5, it is fine-tuned on decompiled code without natural language instructions, therefore, we use the same input form, \ie, decompiled code, as its test input. \looseness=-1

\subsection{Case Study of LLM-generated Summaries}
\label{case:study}

Following the acquisition of statistical results, we perform a more in-depth study into the extent to which LLMs can comprehend binary code, as well as the nuanced interpretation of specific score outcomes.
For this, we perform a series of case studies on LLM-generated summaries. 
Our focus is on samples in which LLMs attain scores at the highest (100th percentile), median (50th percentile) and lowest (0th percentile) semantic similarity scores.
This investigative approach aims to provide us with distributional insights across extreme and median scenarios, thereby shedding light on the robustness of LLM in binary code comprehension.
For source code, our overarching observation is that LLM can capture most of its semantics (as presented in \autoref{tab:source-code-example}).
For example, the summaries of source function 3 that obtain the median score show that the LLM has adeptly discerned the key semantics of ``skip characters'' and ``return the index''.
In addition, we observe a similarly good performance from summaries of decompiled code with debugging symbols (such as samples in \autoref{tab:decomp-debug-example}).
However, for decompiled code from stripped binaries, we find that LLMs cannot generate summaries explicitly matching the semantics of ground-truth summaries (as shown in \autoref{tab:decomp-stripped-example}).
For example, LLMs only manage to capture partial or elementary semantics of functions 2 and 3, despite their scores falling within the median range.
For function 2, the generated summary includes the semantics of ``file name canonicalization'' but it fails to include specific details of the ground truth, such as ``remov2 dots''.
For function 3, LLM only captures the ELF file format and loses all other semantic information.

\finding{Although LLMs excel at source code and decompiled code with symbols, the absence of debugging symbols makes LLMs generate summaries with only partial or elementary semantics of decompiled code.}

For lower-level code, \eg, IR code and assembly code, we observe that LLMs predominantly focus on describing the low-level operations, without providing comprehensive high-level semantic summaries.
For instance, \autoref{tab:ir-code-example} presents the summaries and evaluation scores of the IR code.
Across samples that obtained the highest to lowest scores, LLM-generated summaries mostly describe the operations of data movement, arithmetic calculations (\eg, addition and subtraction), memory loading/writing, and data flow transformation (\eg, jump operation).
Similarly, we also observe such descriptions for assembly code as shown in \autoref{tab:assembly-example}.
Divergent slightly from the IR code summaries, we find that in the case of the highest score for assembly code, LLM can capture essential information related to the string data structure and the copy operation.
\finding{LLM-generated summaries fail to encapsulate the high-level semantics of IR and assembly code, instead, focusing on elucidating the low-level operations, \eg, data movement and arithmetic calculations.}

In the case of raw bytes, one would anticipate that LLMs generate summaries that describe a sequence of bytes. 
However, it is surprising to observe that LLM-generated summaries depict operations akin to those observed in IR code and assembly code (as shown in \autoref{tab:raw-bytes-example}), particularly for OpenAI models, \ie, ChatGPT and GPT-4.
For example, the summary of function 2 describes the stack frame operations, function calls, and arithmetic operations, which are not explicitly manifested in the raw byte input.
It appears that LLMs generate summaries by implicitly lifting the raw bytes into higher-level representations, such as assembly code that can represent the above-mentioned operations.

\finding{The LLM-generated summaries for raw bytes closely resemble those for assembly code, hinting at an implicit process of code lifting, performed by the LLMs.}




In addition to understanding the performance of LLMs in individual samples, we have also confirmed the benefits of using our proposed semantic evaluation metric.
Specificaly, as stated in \S\ref{sub:challenges}, metrics based solely on exact matching fail to provide a precise evaluation outcome, as they cannot encapsulate the semantic nuances of summaries.
For instance, for function 2 in \autoref{tab:source-code-example},  BLEU and METEOR scores are calculated as 0.
However, function 2's generated and ground truth summaries present close semantics, such as ``routine'' and ``function'', as well as ``a filename'' and ``the name of the file''.
These words/phrases are syntactically different, but semantically the same.
Moreover, its calculated scores of BLEU, METETOR, and ROUGE-L are lower than those of function 4, while function 2's summaries have shown closer semantics.
In contrast, our semantic evaluation metric computes semantic similarity at the semantic level, thus it generates fair scores that assign a higher score for closer semantics in function 2 compared to function 4.
Therefore, we argue that our semantic evaluation metric is more suitable for evaluating binary code summaries.

\finding{Our semantic evaluation metric can capture the essential semantics of binary code summaries, which is more suitable for our task than the exact matching-based metrics.}


\input{tables/src-code-examples}

\input{tables/decomp-debug}

\input{tables/decomp-stripped}

\input{tables/ir-code}

\input{tables/assembly}

\input{tables/raw-bytes}

\subsection{Case Study of GPT-4 Results}
\label{sec:gpt-4-result}
\input{tables/chatgpt-vs-gpt4}

In \textbf{RQ2} evaluations, it is surprising to observe that GPT-4 is not the best model, especially when ChatGPT outclasses GPT-4 on the binaries with symbols.
For this, we have conducted a case study to manually investigate the GPT-4 results.
Overall, we observe that GPT-4 occasionally focuses more on noisy details rather than capturing the global semantics of binary functions.
\autoref{tab:gpt4-results} presents three types of such cases, showcasing GPT-4 generated summaries and ground truth along with those generated by ChatGPT for comparison.
For function 1, we find that GPT-4 explains the implementation details of the input binary code, instead of providing a concise summary. 
Such an explanation can capture some semantics, such as its generated phrase ``without changing its current position'' has similar semantics as the ground truth ``preserving the value'', but this explanation includes extra details that do not appear in the ground truth.
Similarly, GPT-4 generates additional details for functions 2 and 3 as highlighted in the red text.
Compared to GPT-4, ChatGPT focuses less on superfluous details and generates more concise summaries that represent the key semantics.

\finding{For binaries with symbols, ChatGPT produces more concise summaries and focuses on essential semantics, showing improvement in brevity and clarity over GPT-4.}

\subsection{BinT5 Outliers and Sample Leakage}
\label{sec:bint5-outlier}

For \textbf{RQ2} evaluation, we have noticed the presence of certain outliers in BinT5's performance, which attain text similarity scores exceeding 0.8. 
These scores are notably higher than the median values.
By comparing the function names of these outlier samples with BinT5's training sets, we find that some of our Ghidra-generated decompiled code samples are in BinT5's training set.
We have removed these leaked samples from our reported results of BinT5.

\subsection{Impact of Symbols and Exclusive Symbol Stripping}
\label{sec:symbol-stripping}
In \textbf{RQ4} evaluations, we study which symbol type (data types, variable names, and function names) contributes the most to binary function semantics by exclusively stripping symbols of each type.
We first treat the decompiled code with all symbols as the original code.
Subsequently, to isolate the impact of each symbol type, we eliminate the respective symbols from the original code by substituting them with non-informative symbols.
These non-informative symbols follows the same pattern as these in decompiler-generated symbols for stripped binaries.
Specifically, for function names, we replace the original names with symbol \texttt{Fun\_addr} in which \texttt{addr} is the address of functions.
For variable names, we replace the original names with symbol \texttt{Var\_idx} where \texttt{idx} is the index of variables in the function.
For data types, we replace the original types with the symbol \texttt{undefined}.

\input{figures/function-name-manipulation}
\input{tables/summary-manipulation}

\subsection{Summary Manipulation by Modifying Function Names}
\label{sec:summary-manipulation}

While we have identified the significant semantic contribution of function names for binary code semantics in \rnumber{9}, we also note an intriguing vulnerability in LLM-generated summaries.
Specifically, we've observed that these summaries can be manipulated by altering function names.
To elaborate, \autoref{fig:summary-manipulation-example} showcases a decompiled function generated by Ghidra, with all symbols removed. 
When we apply our chosen prompt to this function, ChatGPT generates a summary, as demonstrated in row 1 of \autoref{tab:summary-manipulation}.
However, the generated summary undergoes a substantial transformation when we make a simple adjustment: changing the original function name, \texttt{FUN\_0000ea01} into alternative names such as \texttt{quick\_sort}, \texttt{print\_log}, and \texttt{DNS\_flood}. 
As depicted in rows 2 to 4 of \autoref{tab:summary-manipulation}, the resulting binary code summaries closely resemble the semantics of the manipulated function names, but they fail to accurately reflect the true semantics preserved within the function body.

This discovery underscores the susceptibility of LLM-generated summaries to manipulation through function name changes. In light of this vulnerability, adversaries could potentially modify the DWARF entries of malware binaries, thereby misleading LLMs and evading detection of malicious behavior.

\finding{Summaries generated by ChatGPT are susceptible to manipulation via alterations in decompiled code function names.}

\subsection{Other Prompt Engineering Techniques}
\label{sec:other-prompts}

\begin{figure*}[]
\subfloat[Zero-shot Prompting]{\label{fig:zero-shot}{\includegraphics[width=0.33\linewidth]{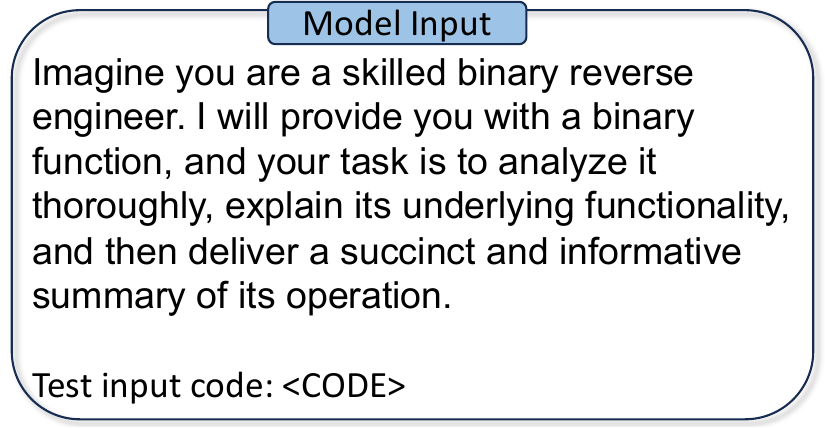}}}
\subfloat[Few-shot Prompting]{\label{fig:few-shot}{\includegraphics[width=0.33\linewidth]{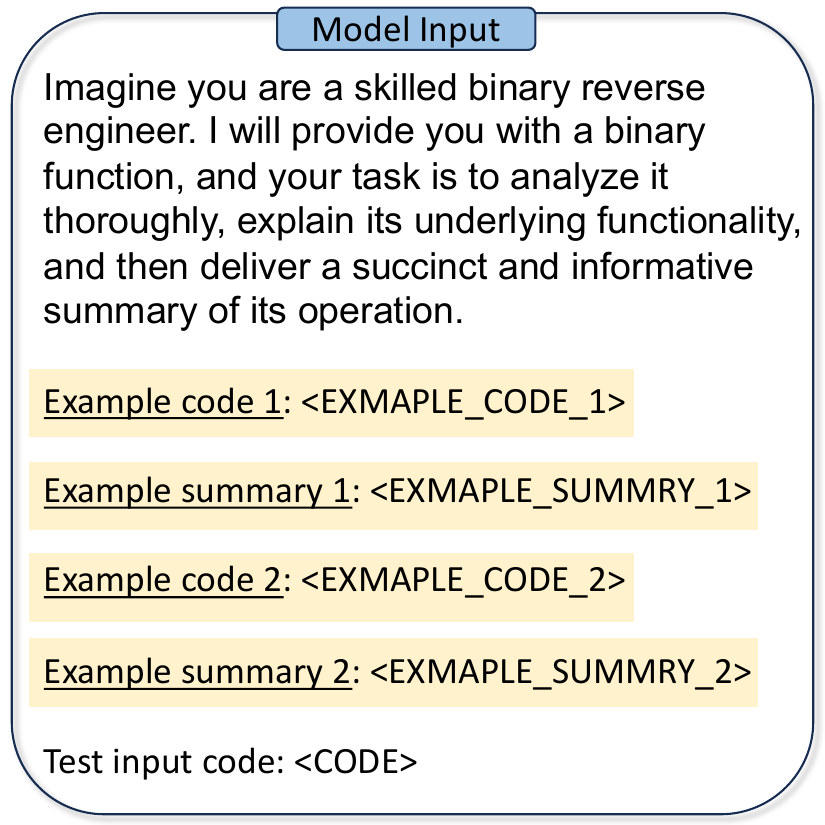}}}
\subfloat[Chain-of-thought Prompting]{\label{fig:cot}{\includegraphics[width=0.33\linewidth]{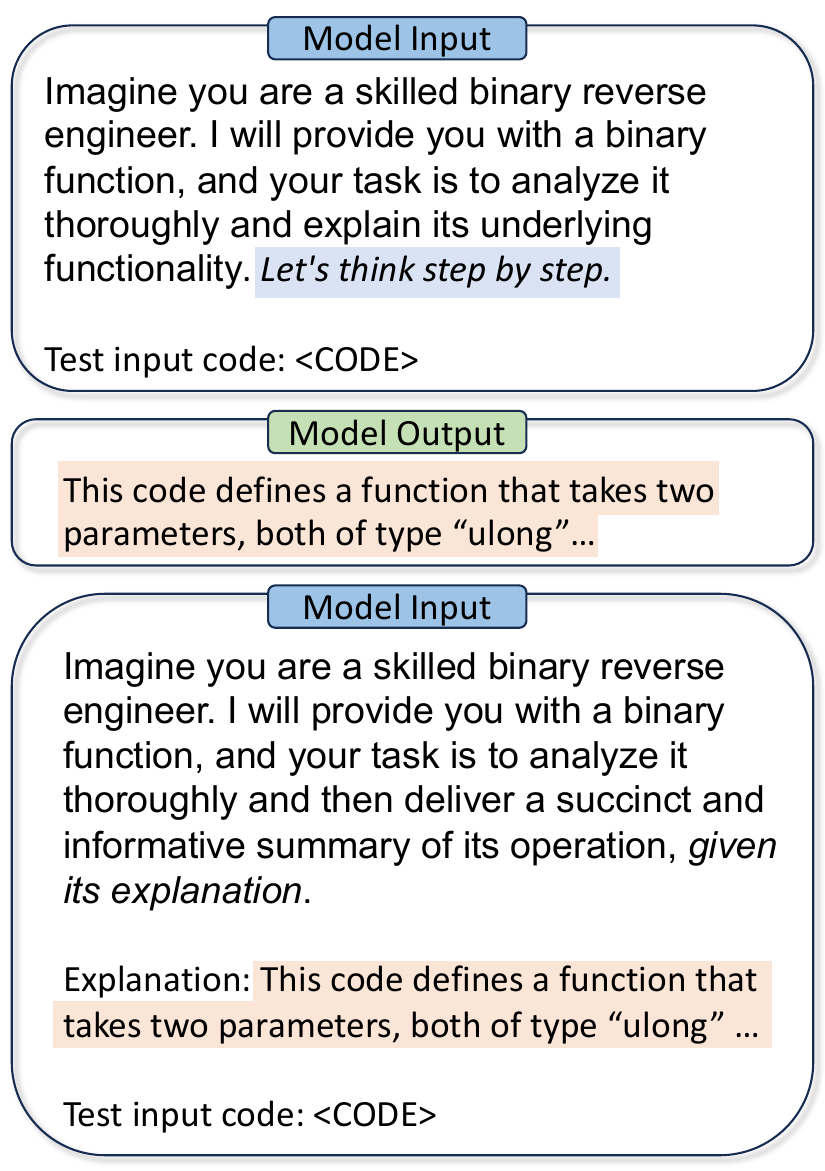}}}
\caption{Example of Zero-shot, Few-shot, and Chain-of-thought Prompting. For few-shot prompting, we add pairs of demonstration examples. For chain-of-thought prompting, we send two requests to LLMs. The first request asks for an explanation of the test input code, which will be parsed and concatenated into the second request, inquiring the code summaries.}
\label{fig:prompt-engineering}
\end{figure*}    

We also examine whether the other prompt engineering techniques can help LLMs understand binary code.
Specifically, we focus on two popular techniques, including few-shot learning and chain-of-thought prompting.
Few-shot prompting was initially introduced to enhance the generalizability of GPT-3 for tasks beyond its primary domain~\cite{brown2020language}. 
This is achieved by imparting the model with contextual knowledge through in-context instructions and demonstration samples.
To be more specific, few-shot learning entails the inclusion of a system instruction $\mathcal{I}$ along with either $n$ in-context demonstration pairs ($\mathcal{X}_{d}$, $\mathcal{Y}_{d}$), where $\mathcal{X}_{d}$ and $\mathcal{Y}_{d}$ represent sample inputs and outputs. This augmentation occurs prior to the introduction of the test input:

\begin{equation}
    x = \{\mathcal{I}; (x_d^{i}, y_d^{i})_{i=1}^{n}; x_{t}\}
\end{equation}
\noindent where the system instruction $\mathcal{I}$ and demonstration samples $(x_d^{i}, y_d^{i})$ are concatenated with the test input $x_{t}$ to form the query input $x$.
\autoref{fig:few-shot} presents the example prompt that we used to perform few-shot prompting.
Instead of using fixed demonstration examples, we randomly sample two pairs of example binary functions and summaries for each test binary code to avoid LLMs learning the unchanged pattern in the demonstration examples.

Chain-of-thought prompting is proposed to resolve complex reasoning problems by breaking them down into intermediate steps~\cite{wei2022chain}.
In the context of binary code understanding, we ask LLMs to perform an additional step of reasoning the input code semantics before generating the final code summaries.
For this, we first send one request asking LLMs to reason about the code semantics, in which we use the popular reasoning prompt ``Let’s think step by step''.
Upon receiving the response, we parse the generated code reasoning results and concatenate them into the second request inquiring about the code summaries.
\autoref{fig:cot} presents the sample of our chat-of-thought prompts.
The baseline is the zero-shot prompt.
As exemplified in ~\autoref{fig:zero-shot}, we directly concatenate the prompt with the test code to generate the request as the baseline approach.


%% file: tables/src-project-lines.tex
\begin{table}[]
\caption{The Source Projects in Our Dataset. The source projects include 11,475,734 lines of code in total.}
\resizebox{0.45\textwidth}{!}{
\begin{tabular}{l|r||l|r}
\toprule
\textbf{Project} & \textbf{\# Code Lines} & \textbf{Project} & \textbf{\# Code Lines} \\ 
\midrule
binutils      & 2,151,781     & gettext      & 1,429,180     \\
libredwg      & 1,073,236     & poke         & 991,146       \\
openssl       & 626,521       & libunistring & 421,487       \\
freeipmi      & 403,180       & coreutils    & 290,065       \\
mailutils     & 274,095       & curl         & 243,426       \\
guile         & 234,008       & gmp          & 188,873       \\
bash          & 177,371       & wget2        & 171,082       \\
findutils     & 168,910       & inetutils    & 166,268       \\
dico          & 153,630       & libiconv     & 148,460       \\
bison         & 140,331       & ncurses      & 138,915       \\
texinfo       & 138,167       & grep         & 136,870       \\
libmicrohttpd & 133,438       & mpfr         & 131,912       \\
diffutils     & 129,150       & tar          & 127,317       \\
lightning     & 112,152       & sed          & 107,964       \\
nettle        & 102,932       & nano         & 94,095        \\
gawk          & 92,484        & libpng       & 87,790        \\
libidn2       & 83,079        & datamash     & 79,427        \\
gzip          & 56,628        & cflow        & 49,403        \\
gss           & 42,415        & readline     & 41,759        \\
direvent      & 41,756        & gama         & 34,949        \\
less          & 31,334        & adns         & 11,820        \\
units         & 9,994         & libtool      & 6,934        \\
\bottomrule
\end{tabular}
}

\label{tab:source-projects}
\end{table}

%% file: tables/binary-function-distribution.tex
\begin{table}[]
\centering
\caption{Binary Functions with Ground-truth Code Summaries across Computer Architectures and Optimization Levels. This dataset includes 557,664 binary functions in total.}
\begin{tabular}{c|c|r}
\toprule

\multicolumn{1}{l|}{\textbf{Architecture}} & \multicolumn{1}{l|}{\textbf{Optimization}} & \textbf{\# Binary Functions} \\
\midrule
\multirow{4}{*}{x64}              & O0                               & 37,304                       \\
                                  & O1                               & 40,476                       \\
                                  & O2                               & 43,543                       \\
                                  & O3                               & 38,263                       \\\midrule
\multirow{4}{*}{x86}              & O0                               & 39,897                       \\
                                  & O1                               & 31,779                       \\
                                  & O2                               & 31,090                       \\
                                  & O3                               & 34,508                       \\\midrule
\multirow{4}{*}{ARM}              & O0                               & 21,649                       \\
                                  & O1                               & 34,287                       \\
                                  & O2                               & 32,641                       \\
                                  & O3                               & 27,453                       \\\midrule
\multirow{4}{*}{MIPS}             & O0                               & 44,745                       \\
                                  & O1                               & 35,283                       \\
                                  & O2                               & 34,589                       \\
                                  & O3                               & 30,158   \\
\bottomrule
\end{tabular}

\label{tab:labeled-binary-functions}
\end{table}

%% file: tables/selected-prompt.tex
\begin{table*}[]
\caption{Evaluation Results of Top Prompts Generated by Our Prompt Synthesis and Optimization Approach}
\resizebox{\textwidth}{!}{
\begin{tabular}{|r|L{0.7\textwidth}|L{0.1\textwidth}|l|l|l|}
\hline
\textbf{ID} & \textbf{Prompt}                                                                                                                                                                                                                                      & \textbf{Semantic Similarity} & \textbf{BLEU} & \textbf{METEOR} & \textbf{ROUGE-L} \\ \hline
1                                 & Imagine you are a skilled binary reverse engineer. I will provide you with a binary function, and your task is to analyze it thoroughly, explain its underlying functionality, and then deliver a succinct and informative summary of its operation. & 0.3723                       & 0.0838        & 0.1138          & 0.1294           \\ \hline
2                                 & Please explain the purpose and functionality of the code.                                                                                                                                                                                            & 0.3719                       & 0.0802        & 0.1041          & 0.1425           \\ \hline
3                                 & Give me an in-depth overview of how it operates and summarize its functionality.                                                                                                                                                                     & 0.3709                       & 0.0812        & 0.1074          & 0.1366           \\ \hline
4                                 & I'd like you to assume the role of a binary reverse engineer. I will supply a binary function, and you will respond by explaining its functionality and providing a concise summary.                                                                 & 0.3701                       & 0.0826        & 0.1069          & 0.1397           \\ \hline
5                                 & Please explain the purpose and usage of function.                                                                                                                                                                                                    & 0.3696                       & 0.0845        & 0.1129          & 0.1396           \\ \hline
6                                 & Please provide a summary of its functionality.                                                                                                                                                                                                       & 0.3694                       & 0.0754        & 0.1004          & 0.1413           \\ \hline
7                                 & Please explain the purpose and behavior of the function/code.                                                                                                                                                                                        & 0.3682                       & 0.0815        & 0.1085          & 0.1423           \\ \hline
8                                 & Please explain how the functions works and what its purpose is.                                                                                                                                                                                      & 0.3667                       & 0.0821        & 0.1111          & 0.1401           \\ \hline
9                                 & As an AI model with expertise in binary code analysis, your objective is to provide a concise and informative summary of the given input code, focusing on the functionality and key aspects of the function.                                        & 0.3665                       & 0.0791        & 0.1063          & 0.1333           \\ \hline
10                                & Please provide a detailed summary of its functionality.                                                                                                                                                                                              & 0.3661                       & 0.0797        & 0.1044          & 0.1402           \\ \hline
11                                & Please explain the purpose and functionality of the input code.                                                                                                                                                                                      & 0.3654                       & 0.0761        & 0.0997          & 0.1397           \\ \hline
12                                & Please describe the purpose or functionality of the given code.                                                                                                                                                                                      & 0.365                        & 0.0759        & 0.1             & 0.1431           \\ \hline
13                                & Please reverse engineer the code and provide a high-level explanation of its functionality.                                                                                                                                                          & 0.364                        & 0.0846        & 0.1082          & 0.1389           \\ \hline
14                                & Give a concise overview of the function's intended purpose.                                                                                                                                                                                          & 0.3639                       & 0.071         & 0.0922          & 0.1422           \\ \hline
15                                & Please define the functionality of the input code.                                                                                                                                                                                                   & 0.363                        & 0.0744        & 0.0985          & 0.143            \\ \hline
16                                & Please provide a detailed explanation of its functionality.                                                                                                                                                                                          & 0.3622                       & 0.0804        & 0.1065          & 0.14             \\ \hline
17                                & Please reverse engineer the input code and provide a detailed explanation of its functionality.                                                                                                                                                      & 0.3622                       & 0.0824        & 0.1067          & 0.1366           \\ \hline
18                                & Please provide a summary of the function's purpose.                                                                                                                                                                                                  & 0.3615                       & 0.0718        & 0.0938          & 0.1381           \\ \hline
19                                & Reverse engineer the code and provide a high-level explanation of its functionality.                                                                                                                                                                 & 0.3612                       & 0.084         & 0.1091          & 0.1386           \\ \hline
20                                & Seeking help from skilled reverse engineers: I'll supply a binary function, and you're requested to provide a brief summary of its purpose.                                                                                                          & 0.3608                       & 0.0836        & 0.1035          & 0.1352           \\ \hline
21                                & I want you to act as a binary reverse engineer. I will provide binary function and you will reply with what this function does and give a brief summary.                                                                                             & 0.3605                       & 0.075         & 0.1002          & 0.1364           \\ \hline
22                                & As an AI assistant specialized in binary code comprehension, you are provided with a binary function code. Your task is to succinctly summarize it in a human-readable form.                                                                         & 0.3604                       & 0.0731        & 0.0964          & 0.1367           \\ \hline
23                                & You are an AI model trained to understand binary code. The task for you is to summarize the input code of a function.                                                                                                                                & 0.3589                       & 0.0722        & 0.0951          & 0.1356           \\ \hline
24                                & Please determine the purpose or functionality of the given code.                                                                                                                                                                                     & 0.3584                       & 0.0718        & 0.0934          & 0.142            \\ \hline
25                                & In your role as an AI assistant proficient in binary code comprehension, you're presented with a binary function code. Could you please provide a human-readable summary of it?                                                                      & 0.3565                       & 0.0755        & 0.1026          & 0.1329           \\ \hline
26                                & You possess expertise in binary code comprehension, and your objective is to condense the input code of a function into a summary.                                                                                                                   & 0.3542                       & 0.0634        & 0.082           & 0.1261           \\ \hline
27                                & As a helpful AI assistant for binary code understanding, you are given a binary function code, Can you summarize it in human language?                                                                                                               & 0.3528                       & 0.0759        & 0.1014          & 0.1373           \\ \hline
28                                & Calling out for assistance from knowledgeable reverse engineers. I will provide a binary function and you will summarize it into a short sentence.                                                                                                   & 0.3528                       & 0.0691        & 0.0894          & 0.1307           \\ \hline
29                                & Could you offer a brief description of how it functions?                                                                                                                                                                                             & 0.3523                       & 0.0713        & 0.094           & 0.1347           \\ \hline
30                                & Hello ChatGPT. You are about to immerse yourself into the role of code understanding machine. Generate a short summary of the following binary function.                                                                                             & 0.3521                       & 0.0824        & 0.1117          & 0.131            \\ \hline
31                                & I'd like you to act as a reverse engineer who can summarize a user-input binary function into a concise sentence.                                                                                                                                    & 0.3519                       & 0.0776        & 0.0971          & 0.1347           \\ \hline
32                                & Please explain how the input function work.                                                                                                                                                                                                          & 0.3518                       & 0.0829        & 0.1105          & 0.1359           \\ \hline
33                                & Generate a succinct summary for the given binary function.                                                                                                                                                                                           & 0.351                        & 0.0788        & 0.1028          & 0.1344           \\ \hline
34                                & Produce a concise summary for the provided binary function.                                                                                                                                                                                          & 0.3509                       & 0.0743        & 0.0963          & 0.1354           \\ \hline
35                                & Please acknowledge my following request. I will ask for a short summary of the following binary function.                                                                                                                                            & 0.3488                       & 0.0754        & 0.0991          & 0.1291           \\ \hline
36                                & Come up with a concise summary of the following binary function.                                                                                                                                                                                     & 0.3476                       & 0.0765        & 0.1004          & 0.1325           \\ \hline
37                                & Act as a reverse engineer that takes user-input binary function and summarize it into a short sentence.                                                                                                                                              & 0.3476                       & 0.0663        & 0.085           & 0.1315           \\ \hline
38                                & Reverse engineer the provided binary function and distill its essence into a concise one-sentence summary.                                                                                                                                           & 0.346                        & 0.0803        & 0.1036          & 0.1344           \\ \hline
39                                & Assume that you are a reverse engineer that takes user-input binary function. Summarize it into a short sentence.                                                                                                                                    & 0.3398                       & 0.0692        & 0.0905          & 0.1298           \\ \hline
40                                & Function as a reverse engineer who can encapsulate a user-input binary function in a concise sentence.                                                                                                                                               & 0.3296                       & 0.0693        & 0.0879          & 0.123            \\ \hline
\end{tabular}
}
\label{tab:top-prompts}
\end{table*}

%% file: tables/src-code-examples.tex
\begin{table*}[]
\caption{Examples of LLM-generated Summaries for Source Code and Evaluation Scores. \bfline{Semantic overlaps} are highlighted. We investigated sample summaries by the highest (100th percentile), median (50th percentile), and lowest (0th percentile) text similarity scores. }
\resizebox{\textwidth}{!}{
\begin{tabular}{|l|L{0.08\textwidth}|L{0.3\textwidth}|L{0.3\textwidth}|C{0.1\textwidth}|c|c|c|}
\hline
\textbf{ID} & \textbf{Score Range} & \textbf{Groud Truth Summary}                                                                                                                                     & \textbf{LLM-generated Summary}                                                                                                                                                                                       & \textbf{Semantic Similarity} & \textbf{BLEU} & \textbf{METEOR} & \textbf{ROUGE-L} \\ \hline
1           & Highest                    & This function \textbf{\underline{frees all memory}} allocated for given \textbf{\underline{hash table}}. Naturally the hash table must \textbf{\underline{already exist}}. & The function deletes a \bfline{hash table}. It iterates over the table entries, applies a delete function \bfline{if it exists}, and then \bfline{frees the memory} for the entries and the table itself using the appropriate free function.   & 0.833                    & 0.222         & 0.212           & 0.226            \\ \hline
2           & \multirow{2}{*}{Median}    & \bfline{Internal routine} to \bfline{get a filename} from open\_files  & \bfline{The function} returns \bfline{the name of} \bfline{the file} associated with the given file pointer. It checks if the file pointer matches with standard input, output, or error, otherwise returns "\textless{}unknown\textgreater{}". & 0.483                    & 0.000         & 0.000           & 0.051            \\ \cline{1-1} \cline{3-8} 
3           &                           & \bfline{Skip characters} until the index becomes greater than NEW\_RAW\_IDX. \bfline{Return the index}.                       & The function advances through a string, \bfline{skipping characters} until a specified index. It handles multibyte characters and updates the last wide character processed. It \bfline{returns the} \bfline{new raw buffer index}.             & 0.483                    & 0.167         & 0.348           & 0.318            \\ \hline
4           & Lowest                      & \bfline{Insert} a BFD into the cache.                                                                                & The function \bfline{inserts} an object into a doubly linked list. If the list is empty, it creates a new list with the object as the only element. Otherwise, it adds the object to the end of the list.                     & 0.011                    & 0.079         & 0.217           & 0.136            \\ \hline
\end{tabular}
}
\label{tab:source-code-example}
\end{table*}

%% file: tables/decomp-debug.tex
\begin{table*}[]
\caption{Examples of LLM-generated Summaries for Decompiled Code with Debugging Symbols and Evaluation Scores. \bfline{Semantic overlaps} are highlighted. LLMs can capture sufficient semantics of samples with the best and median text similarity scores. }
\resizebox{\textwidth}{!}{
\begin{tabular}{|l|L{0.08\textwidth}|L{0.3\textwidth}|L{0.3\textwidth}|C{0.1\textwidth}|c|c|c|}
\hline
\textbf{ID} & \textbf{Score Range} & \textbf{Groud Truth Summary}                                                                                                                                     & \textbf{LLM-generated Summary}                                                                                                                                                                                       & \textbf{Semantic Similarity} & \textbf{BLEU} & \textbf{METEOR} & \textbf{ROUGE-L} \\ \hline
1                                 & Highest                    & This function \bfline{clears a specified slot} in a \bfline{hash table}. It is useful when you've already done the lookup and don't want to do it again. & The function \bfline{clears a slot} in a \bfline{hash table}. If the slot is within the table's range and its value is greater than a constant, it deletes the slot's content, replaces it with a constant, and increments the deletion counter. & 0.864                    & 0.321         & 0.308           & 0.226            \\ \hline
2                                 & \multirow{2}{*}{Median}    & \bfline{Allocate N bytes of memory} dynamically, with \bfline{error checking}.                                                                           & The function \bfline{allocates memory of} \bfline{size 'n'} using malloc. If allocation fails, it attempts to fix \bfline{the null allocation} before returning the pointer.                                                                              & 0.465                    & 0.087         & 0.144           & 0.125            \\ \cline{1-1} \cline{3-8} 
3                                 &                           & \bfline{Get the contents of a section} in an \bfline{Intel Hex file}.  & The function \bfline{reads a section} from an \bfline{Intel HEX file}, allocates memory for it, and copies the contents to a specified location. It handles errors and performs stack check failure.                                             & 0.465                    & 0.167         & 0.492           & 0.341            \\ \hline
4                                 & Lowest                      & Find the first occurrence of NEEDLE in HAYSTACK.                                                                                       & The function halts the program execution when it encounters bad instruction data, truncating the control flow.                                                                                                                 & -0.038                   & 0.063         & 0.057           & 0.083            \\ \hline
\end{tabular}
}
\label{tab:decomp-debug-example}
\end{table*}

%% file: tables/decomp-stripped.tex
\begin{table*}[]
\caption{Examples of LLM-generated Summaries for Decompiled Code from Stripped Binaries and Evaluation Scores. \bfline{Semantic overlaps} are highlighted. Unlike source code and decompiled code with debugging symbols, LLM-generated summaries for decompiled code from stripped binaries cannot explicitly match ground-truth summaries. }
\resizebox{\textwidth}{!}{
\begin{tabular}{|l|L{0.08\textwidth}|L{0.3\textwidth}|L{0.3\textwidth}|C{0.1\textwidth}|c|c|c|}
\hline
\textbf{ID} & \textbf{Score Range} & \textbf{Groud Truth Summary}                                                                                                                                     & \textbf{LLM-generated Summary}                                                                                                                                                                                       & \textbf{Semantic Similarity} & \textbf{BLEU} & \textbf{METEOR} & \textbf{ROUGE-L} \\ \hline
1                                 & 100-th                    & \bfline{Copy} private \bfline{section} information. This \bfline{copies over the entsize field}, and sometimes the info field.                                                                     & The function checks if certain conditions are met for parameters at \bfline{specific memory locations}. If true, it \bfline{copies data from one location to} \bfline{another} and calls another function. & 0.763                    & 0.071         & 0.092           & 0.093            \\ \hline
2                                 & \multirow{2}{*}{50-th}    &  Return the \bfline{canonical} absolute name of \bfline{file NAME}.  A canonical name
   does not contain any ".", ".." components nor any repeated file name
   separators ('/') or symlinks.  All components must exist.
   The result is malloc'd.     & The function takes a \bfline{file name} as input and attempts to \bfline{canonicalize} it, but encounters bad instruction data and halts execution.  & 0.243                    & 0.073         & 0.111           & 0.148            \\ \cline{1-1} \cline{3-8} 
3                                 &                           &  Set up an \bfline{ELF} internal section header for a section.     & The function checks and manipulates various parameters and flags related to \bfline{ELF} (Executable and Linkable Format) files. It performs operations like string length calculation, memory copying, error and warning handling, and calls other functions for further processing.                            & 0.183                    & 0.045         & 0.049           & 0.065            \\ \hline
4                                 & 0-th                      & Return the number of buckets in the hash table.  The table size, the total number of buckets (used plus unused), or the maximum number of slots, are the same quantity. & The function returns the value at the memory location offset by 0x10 from the input long integer parameter.                                                                     & -0.141                   & 0.114         & 0.087           & 0.167            \\ \hline
\end{tabular}
}
\label{tab:decomp-stripped-example}
\end{table*}

%% file: tables/ir-code.tex
\begin{table*}[]
\caption{Examples of LLM-generated Summaries for IR Code and Evaluation Scores. \bfline{Semantic overlaps} are highlighted. The generated summaries for IR code mostly describe low-level operations. \eg, data movement and arithmetic operations.}
\resizebox{\textwidth}{!}{
\begin{tabular}{|l|L{0.08\textwidth}|L{0.3\textwidth}|L{0.3\textwidth}|C{0.1\textwidth}|c|c|c|}
\hline
\textbf{ID} & \textbf{Score Range} & \textbf{Groud Truth Summary}                                                                                                                                     & \textbf{LLM-generated Summary}                                                                                                                                                                                       & \textbf{Semantic Similarity} & \textbf{BLEU} & \textbf{METEOR} & \textbf{ROUGE-L} \\ \hline
1                                 & Highest                    & \bfline{This function} needs to check if the ciphers required are actually available     & \bfline{The function} performs various operations including moving, subtracting, loading, extending, shifting, setting, adding, and or-ing values. It also calls the 'ssl3\_get\_cipher\_by\_id' function. & 0.361                    & 0.091         & 0.068           & 0.154            \\ \hline
2                                 & \multirow{2}{*}{Median}    & Complain about a programming error and exit.                                    & The function performs a series of data manipulations and arithmetic operations, including addition, subtraction, and bitwise operations. It also includes conditional jumps, function calls, and memory load/store operations. The function ends with an indirect jump. & 0.106                    & 0.029         & 0.051           & 0.047            \\ \cline{1-1} \cline{3-8} 
3                                 &                           & Initializes the iterator, but one has to call next to get to the first element. & The function performs a series of data movements, arithmetic operations, and memory store/load operations. It also includes a non-operational instruction (nop) and an indirect jump (ijmp).                                                                            & 0.098                    & 0.000         & 0.029           & 0.047            \\ \hline
4                                 & Lowest                      & Create a new index list containing only a given index.                          & The function allocates memory for two integers, initializes them to 1, and then stores them in a dynamically allocated array. It then retrieves the values, performs some arithmetic operations, and finally returns the results.                                       & -0.018                   & 0.029         & 0.040           & 0.045            \\ \hline
\end{tabular}
}
\label{tab:ir-code-example}
\end{table*}

%% file: tables/assembly.tex
\begin{table*}[]
\caption{Examples of LLM-generated Summaries for Assembly Code and Evaluation Scores. \bfline{Semantic overlaps} are highlighted. While the generated summaries mostly describe low-level operations, LLMs can sometimes capture high-level data structures and their properties, such as strings and length, in assembly code.}
\resizebox{\textwidth}{!}{
\begin{tabular}{|l|L{0.08\textwidth}|L{0.3\textwidth}|L{0.3\textwidth}|C{0.1\textwidth}|c|c|c|}
\hline
\textbf{ID} & \textbf{Score Range} & \textbf{Groud Truth Summary}                                                                                                                                     & \textbf{LLM-generated Summary}                                                                                                                                                                                       & \textbf{Semantic Similarity} & \textbf{BLEU} & \textbf{METEOR} & \textbf{ROUGE-L} \\ \hline
1                                 & Highest                    & Return the \bfline{length} (in bytes) of the character at the start of the  * given \bfline{string}, and return a \bfline{copy} of this character in *thechar.                                       & The function \bfline{copies} a \bfline{string} from source to destination, with a \bfline{length} determined by a function call, and handles potential errors.                                      & 0.580                    & 0.118         & 0.102           & 0.133            \\ \hline
2                                 & \multirow{2}{*}{Median}    & Return a string describing the parameter type.                                                                                                                            & The function compares a value with 0x75, performs a jump if above, calculates an address using a scaled index, adds to a register, and returns various memory addresses. & 0.113                    & 0.042         & 0.072           & 0.067            \\ \cline{1-1} \cline{3-8} 
3                                 &                           & Print the first N statements in statement list S to STDERR.  If N == 0, nothing is printed. If N \textless 0, the entire list is printed. Intended to be called from GDB. & The function performs a series of data movements, comparisons, and conditional jumps. It also calls another function at 0x0003f700 twice, possibly for data processing.  & 0.101                    & 0.000         & 0.031           & 0.073            \\ \hline
4                                 & Lowest                      & Add 'pattern' to the list of patterns for which files that match are not listed.                                                                                          & The function saves a value, performs a call operation, updates memory locations with specific register values, and then restores the initial value.                      & -0.145                   & 0.045         & 0.032           & 0.054            \\ \hline

\end{tabular}
}
\label{tab:assembly-example}
\end{table*}

%% file: tables/raw-bytes.tex
\begin{table*}[]
\caption{Examples of LLM-generated Summaries for Raw Bytes and Evaluation Scores. \bfline{Semantic overlaps} are highlighted. Compared to assembly code summaries, the summaries of raw bytes exhibit similar operational descriptions, which exhibit the implicit code-lifting process by LLMs.}
\resizebox{\textwidth}{!}{
\begin{tabular}{|l|L{0.08\textwidth}|L{0.3\textwidth}|L{0.3\textwidth}|C{0.1\textwidth}|c|c|c|}
\hline
\textbf{ID} & \textbf{Score Range} & \textbf{Groud Truth Summary}                                                                                                                                     & \textbf{LLM-generated Summary}                                                                                                                                                                                       & \textbf{Semantic Similarity} & \textbf{BLEU} & \textbf{METEOR} & \textbf{ROUGE-L} \\ \hline
1                                 & Highest                    & Return 1 if \bfline{STRING} contains an \bfline{absolute pathname}, else 0.  Used by `cd' to decide whether or not to look up a directory name in \$CDPATH. & The function checks if the \bfline{input} \bfline{string starts with a period or slash}, and if the second character (if exists) is also a period or slash.                                             & 0.592                    & 0.161         & 0.307           & 0.217            \\ \hline
2                                 & \multirow{2}{*}{Median}    & Determine if s2 occurs in s1.  If so, return a pointer to the match in s1.  The compare is case insensitive.                              & The function sets up a stack frame, performs two function calls, executes a loop with conditional jumps and arithmetic operations, and then cleans up the stack frame.                & 0.113                    & 0.111         & 0.069           & 0.083            \\ \cline{1-1} \cline{3-8} 
3                                 &                           & Free any storage consumed by PARSER (but not PARSER itself).                                                                              & The function performs various operations on memory addresses, including comparisons, conditional jumps, and function calls. It also manipulates and checks specific bits in the data. & 0.104                    & 0.000         & 0.000           & 0.000            \\ \hline
4                                 & Lowest                      & Add a volume label to a part of multi-volume archive                                                                                      & The function performs several memory operations, calls to other functions, and arithmetic operations. It also includes error checking and stack frame management.                     & -0.095                   & 0.045         & 0.089           & 0.061            \\ \hline
\end{tabular}
}
\label{tab:raw-bytes-example}
\end{table*}

%% file: tables/chatgpt-vs-gpt4.tex
\begin{table*}[]
\caption{Examples of Summaries Generated by GPT-4 and ChatGPT for Binaries with Symbols. The \textcolor{red}{red text} highlights the noisy details that do not appear in the ground truth summaries, and \bfline{semantic overlaps} are also highlighted.}
\resizebox{\textwidth}{!}{
\begin{tabular}{|L{0.02\textwidth}|L{0.3\textwidth}|L{0.3\textwidth}|L{0.3\textwidth}|R{0.05\textwidth}|R{0.05\textwidth}|}
\hline
\multirow{2}{*}{\textbf{ID}} & \multirow{2}{*}{\textbf{Groud Truth Summary}}                                                                                               & \multicolumn{2}{c|}{\textbf{LLM-generated Summary}}                                                                                                                                                                                                                                                                                                            & \multicolumn{2}{c|}{\textbf{Semantic Similarity}}                           \\ \cline{3-6} 
                             &                                                                                                                                             & \multicolumn{1}{c|}{\textbf{GPT-4}}                                                                                                                                                                            & \multicolumn{1}{c|}{\textbf{ChatGPT}}                                                                                                         & \multicolumn{1}{c|}{\textbf{GPT-4}} & \multicolumn{1}{c|}{\textbf{ChatGPT}} \\ \hline
1          & \bfline{Clear} the stream's \bfline{ungetc buffer}, \bfline{preserving} the value of ftello (fp).                                                                      & \textcolor{red}{Checks if a specific flag is set for a file pointer. If the flag is set, it repositions the file pointer without changing its current position.}                                           & \bfline{Clears the ungetc buffer} of a file while \bfline{preserving} the current position if the file has the \_IO\_UNGETC flag set.                           & 0.375                           & 0.647                           \\ \hline
2          & \bfline{Return quoting options} for \bfline{STYLE}, with no extra quoting.                                                                                    & Initializes a \bfline{quoting\_options} structure based on the provided \bfline{quoting\_style}. \textcolor{red}{If the style is 'custom\_quoting\_style', it aborts. It also performs stack protection checks.}             & \bfline{Returns} a structure with \bfline{quoting options} based on the input \bfline{quoting style}.                                                                    & 0.457                           & 0.637                           \\ \hline
3          & \bfline{Set program\_name}, based on \bfline{argv{[}0{]}}. argv0 must be a \bfline{string} allocated with indefinite extent, and must not be modified after this call. & \bfline{Sets the program name} based on the input \bfline{string}. \textcolor{red}{It checks for specific substrings and adjusts the program name accordingly. If the input is null, it writes an error message and aborts.} & \bfline{Set program name} based on the \bfline{input argument}, \textcolor{red}{handling different cases and conditions. Print error message and abort if the argument is NULL.} & 0.536                           & 0.611                           \\ \hline
\end{tabular}
}
\label{tab:gpt4-results}
\end{table*}

%% file: figures/function-name-manipulation.tex
\begin{figure}[t]
\centering
\begin{minted}[mathescape,
               linenos,
               numbersep=5pt,
               %gobble=2,
               % frame=lines,
               frame=single,
               fontsize=\scriptsize,
               %fontsize=\footnotesize,
               %framesep=2mm,
               xleftmargin=10pt
               ]{c}
void FUN_0000ea01(undefined4 param_1,undefined4 param_2,undefined4 param_3,
        undefined4 param_4,undefined4 param_5,undefined4 param_6,
        undefined4 param_7,undefined4 param_8,
        undefined4 param_9,undefined8 param_10,
        undefined8 param_11,undefined8 param_12,
        undefined8 param_13,undefined8 param_14)

{
  undefined4 uVar1;
  char in_AL;
  long in_FS_OFFSET;
  undefined4 local_d8;
  … // skip for simplicity
  DAT_002bd1a8 = uVar1;
  return;
}
\end{minted}
\caption{Ghidra Generated Decompiled Code for \texttt{error\_ln} Function in the Stripped Binary Compiled from the Gawk Project}
\label{fig:summary-manipulation-example}
\end{figure}

%% file: tables/summary-manipulation.tex
\begin{table*}[]
\caption{ChatGPT Generates Different Summaries Given the Same Decompiled Code with Different Function Names. Given the original function name \texttt{FUN\_0000ea01}, ChatGPT generates a summary by analyzing the decompiled function. However, when we change this function name to \texttt{quick\_sort}, \texttt{print\_log}, and \texttt{DNS\_flood}, ChatGPT produces summaries which are closely align with the manipulated function names. The key summary semantics manipulations have been \bfline{highlighted}.}
\resizebox{\textwidth}{!}{
\begin{tabular}
{|l|L{0.2\textwidth}|L{0.8\textwidth}|}
\hline
\textbf{ID} & \textbf{Function Name}             & \textbf{Summary}   \\\hline 
1                    & FUN\_0000ea01 & This function takes multiple parameters, assigns some values to local variables, calls another function, increments a global variable, and prints an error message. \\ \hline
2                                                 & quick\_sort   & \bfline{Sorts an array using the quicksort algorithm}. Displays an error message if an error occurs.                                                                         \\ \hline
3                                                 & print\_log    & \bfline{Print log} with error message and parameters, increment a counter, and handle stack check failure.                                                                   \\ \hline
4                                             & DNS\_flood    & \bfline{DNS flood function} with input parameters, local variables, and a call to the `err` function to print an error message.       \\ \hline
\end{tabular}
}
\label{tab:summary-manipulation}
\end{table*}